%% file: btodll_prd.tex
\documentclass[twocolumn,showpacs,aps,prd,floatfix]{revtex4}

\usepackage{graphicx}
\usepackage{dcolumn}
\usepackage{amsmath}
\usepackage{epsfig}
\usepackage{multirow}
\usepackage{color}

\input babarsym
\include{btodllsym}

\newcommand{\SLACPubNumber} {15401}

\def\figurebox#1#2#3{%
    \def\arg{#3}%
    \ifx\arg\empty
    {\hfill\vbox{\hsize#2\hrule\hbox to #2{\vrule\hfill\vbox to #1{\hsize#2\vfill}\vrule}\hrule}\hfill}%
    \else
    {\hfill\epsfbox{#3}\hfill}%
    \fi}

\begin{document}

\preprint{\babar-PUB-12/33} 
\preprint{SLAC-PUB-\SLACPubNumber} 
%\preprint{SLAC-PUB-15401} 

\begin{flushleft}
%\babar\ Analysis Document \#2507; Version 6\\
\babar-PUB-12/33\\
SLAC-PUB-\SLACPubNumber\\
%SLAC-PUB-15401\\
%hep-ex/\LANLNumber\\[10mm]
\end{flushleft}

\title{
  {\large \bf
    A Search for the Rare Decays \B\to\pill\ and \Bz\to\etall 
  }
}

% Dummy author list; contact PubBoard Chair for current author list
\input authors_dec2012

%\date{\today}% It is always \today, today, but you may specify any date with \date
\date{March 20, 2013}% It is always \today, today, but you may specify any date with \date.

\begin{abstract}
  %We  present the  results of  a search  for the  rare flavor-changing
  %neutral-current    decays    \B\to\pill\    ($\pi=\pipm,\piz$    and
  %$\ell=e,\mu$) and     \Bz\to\etall      using     a     sample     of
  %$\epem\to\FourS\to\BB$   decays   corresponding   to  428\invfb   of
  %integrated  luminosity  collected   by  the  \babar\  detector.   No
  %significant signal  is observed,  and we set  an upper limit  on the
  %isospin   and   lepton-flavor   averaged   branching   fraction   of
  %$\br(\B\to\pill)<7.0\times  10^{-8}$  and  a lepton-flavor  averaged
  %upper  limit of $\br(\Bz\to\etall)<9.2\times  10^{-8}$, both  at the
  %90\% confidence level.  We also report 90\% confidence level branching
  %fraction upper limits for the individual modes \BtoPiEE, \BtoPiZEE,
  %\BtoPiMuMu, \BtoPiZMuMu, \BtoEtaEE, and \BtoEtaMuMu.
\input{abstract}
\end{abstract}

%\pacs{13.25.Hw, 12.15.Hh, 11.30.Er}% PACS, the Physics and Astronomy Classification Scheme.
% PACS, the Physics and Astronomy Classification Scheme.

\maketitle

\input{introduction}
\input{expsetup_dataset}
\input{evtreco_candselec}
\input{fit_model}
\input{nn_cut_opt}
\input{fit_validation}

\input{systematics}

\input{results}
% Input the pubboard acknowledgements file
\section{Acknowledgements}
\input acknowledgements.tex

%\bibliographystyle{prsty}
%%\bibliographystyle{apsrev4-1}
%\bibliography{btodll_prd_linenum}

\end{document}

%% file: btodllsym.tex
\def\BtoPiEE          {\ensuremath{B^+ \to \pi^+ e^+ e^-}\xspace}
\def\BtoPiZEE         {\ensuremath{B^0 \to \pi^0 e^+ e^-}\xspace}

\def\BtoEtaEE         {\ensuremath{B^0 \to \eta e^+ e^-}\xspace}

\def\BtoEtaGGEE       {\ensuremath{B^0 \to \eta_{\gamma \gamma} e^+ e^-}\xspace}
\def\BtoEtaThPiEE     {\ensuremath{B^0 \to \eta_{3\pi} e^+ e^-}\xspace}
\def\BtoPiMuMu        {\ensuremath{B^+ \to \pi^+ \mu^+ \mu^-}\xspace}
\def\BtoPiZMuMu       {\ensuremath{B^0 \to \pi^0 \mu^+ \mu^-}\xspace}

\def\BtoEtaMuMu       {\ensuremath{B^0 \to \eta \mu^+ \mu^-}\xspace}
\def\BtoEtaGGMuMu     {\ensuremath{B^0 \to \eta_{\gamma \gamma} \mu^+ \mu^-}\xspace}
\def\BtoEtaThPiMuMu   {\ensuremath{B^0 \to \eta_{3\pi} \mu^+ \mu^-}\xspace}

\def\rhopll                    {\ensuremath{\rho^+\ellell}\xspace}
\def\rhozll                    {\ensuremath{\rho^0\ellell}\xspace}

\def\etall                    {\ensuremath{\eta\ellell}\xspace}

\def\etaee         {\ensuremath{\eta \epem}}
\def\etaggee       {\ensuremath{\eta_{\gamma\gamma} \epem}}
\def\etathpiee     {\ensuremath{\eta_{3\pi} \epem}}
\def\etamm         {\ensuremath{\eta \mumu}}
\def\etaggmm       {\ensuremath{\eta_{\gamma\gamma} \mumu}}
\def\etathpimm     {\ensuremath{\eta_{3\pi} \mumu}}   

\def\Kll              {\ensuremath{K^+\ellell}\xspace}
\def\Kee              {\ensuremath{K^+ \epem}}
\def\Kmm              {\ensuremath{K^+ \mumu}}
\def\KSll          {\ensuremath{K^0_{\scriptstyle S}\ellell}\xspace}

\def\Kstarll          {\ensuremath{K^*\ellell}\xspace}

\def\BtoPiEE          {\ensuremath{B^+ \to \pi^+ e^+ e^-}\xspace}
\def\BtoPiZEE         {\ensuremath{B^0 \to \pi^0 e^+ e^-}\xspace}

\def\BtoEtaEE         {\ensuremath{B^0 \to \eta e^+ e^-}\xspace}

\def\BtoEtaGGEE       {\ensuremath{B^0 \to \eta_{\gamma \gamma} e^+ e^-}\xspace}
\def\BtoEtaThPiEE     {\ensuremath{B^0 \to \eta_{3\pi} e^+ e^-}\xspace}
\def\BtoPiMuMu        {\ensuremath{B^+ \to \pi^+ \mu^+ \mu^-}\xspace}
\def\BtoPiZMuMu       {\ensuremath{B^0 \to \pi^0 \mu^+ \mu^-}\xspace}

\def\BtoEtaMuMu       {\ensuremath{B^0 \to \eta \mu^+ \mu^-}\xspace}
\def\BtoEtaGGMuMu     {\ensuremath{B^0 \to \eta_{\gamma \gamma} \mu^+ \mu^-}\xspace}
\def\BtoEtaThPiMuMu   {\ensuremath{B^0 \to \eta_{3\pi} \mu^+ \mu^-}\xspace}

\def\rhopll                    {\ensuremath{\rho^+\ellell}\xspace}
\def\rhozll                    {\ensuremath{\rho^0\ellell}\xspace}

\def\etall                    {\ensuremath{\eta\ellell}\xspace}

\def\Kll              {\ensuremath{K^+\ellell}\xspace}
\def\Kee              {\ensuremath{K^+ \epem}}
\def\Kmm              {\ensuremath{K^+ \mumu}}
\def\KSll          {\ensuremath{K^0_{\scriptstyle S}\ellell}\xspace}

\def\Kstarll          {\ensuremath{K^*\ellell}\xspace}

\def\ie{{\em i.e.,}}
\def\eg{{\em e.g.,}}

\def\etapr      {\ensuremath{\eta^{\prime}}\xspace}

\newcommand{\multc}[3]{\multicolumn{#1}{#2}{#3}}

\def\mll           {\ensuremath{m_{\ell\ell}}}

\def\etagg         {\ensuremath{\eta_{\gamma\gamma}}}
\def\etathreepi    {\ensuremath{\eta_{3\pi}}}
\def\rhop          {\ensuremath{\rho^+}}

\def\rhoz          {\ensuremath{\rho^0}}

\def\btodll        {\ensuremath{b \to d \ellell}\xspace}
\def\Btopill       {\ensuremath{B \to \pi \ellell}\xspace}

\def\threepi       {\ensuremath{\pi^+\pi^-\pi^0}}

\def\pill          {\ensuremath{\pi \ellell}}
\def\pipll         {\ensuremath{\pi^+ \ellell}}
\def\pizll         {\ensuremath{\pi^0 \ellell}}
\def\piee          {\ensuremath{\pi \epem}}
\def\pipee         {\ensuremath{\pi^+ \epem}}
\def\pizee         {\ensuremath{\pi^0 \epem}}
\def\pimm          {\ensuremath{\pi \mumu}}
\def\pipmm         {\ensuremath{\pi^+ \mumu}}
\def\pizmm         {\ensuremath{\pi^0 \mumu}}

%% file: authors_dec2012.tex
%% author list as of 09-Dec-2012 (339 authors)
%
\author{J.~P.~Lees}
\author{V.~Poireau}
\author{V.~Tisserand}
\affiliation{Laboratoire d'Annecy-le-Vieux de Physique des Particules (LAPP), Universit\'e de Savoie, CNRS/IN2P3,  F-74941 Annecy-Le-Vieux, France}
\author{E.~Grauges}
\affiliation{Universitat de Barcelona, Facultat de Fisica, Departament ECM, E-08028 Barcelona, Spain }
\author{A.~Palano$^{ab}$ }
\affiliation{INFN Sezione di Bari$^{a}$; Dipartimento di Fisica, Universit\`a di Bari$^{b}$, I-70126 Bari, Italy }
\author{G.~Eigen}
\author{B.~Stugu}
\affiliation{University of Bergen, Institute of Physics, N-5007 Bergen, Norway }
\author{D.~N.~Brown}
\author{L.~T.~Kerth}
\author{Yu.~G.~Kolomensky}
\author{G.~Lynch}
\affiliation{Lawrence Berkeley National Laboratory and University of California, Berkeley, California 94720, USA }
% ...
\collaboration{The \babar\ Collaboration}
\noaffiliation

%% file: abstract.tex
We  present the  results of  a search  for the  rare flavor-changing
neutral-current    decays    $B\rightarrow\pi\ell^+\ell^-$    ($\pi=\pi^{\pm},\piz$    and
$\ell=e,\mu$) and  $B^0\rightarrow\eta\ell^+\ell^-$      using     a     sample     of
$e^+e^-\rightarrow\Upsilon(4S)\rightarrow B \kern 0.18em\overline{\kern -0.18em B}{}$   
decays   corresponding   to  428 $\mbox{\,fb}^{-1}$   of
integrated  luminosity  collected   by  the  
$\mbox{\slshape B\kern-0.1em{\smaller A}\kern-0.1em B\kern-0.1em{\smaller A\kern-0.2em R}}$  detector.   No
significant signal  is observed,  and we set  an upper limit  on the
isospin   and   lepton-flavor   averaged   branching   fraction   of
$\cal{B}(B\rightarrow\pi\ell^+\ell^-){\rm < 7.0}$$\times  10^{-8}$  and  a lepton-flavor  averaged
upper  limit of $\cal{B}( B^{\rm 0} \rightarrow\eta\ell^+\ell^-){\rm < 9.2}$$\times  10^{-8}$, both  at the
90\% confidence level.  We also report 90\% confidence level branching
fraction upper limits for the individual modes $B^+ \rightarrow \pi^+ e^+ e^-$, $B^0 \rightarrow \pi^0 e^+ e^-$,
$B^+ \rightarrow \pi^+ \mu^+ \mu^-$, $B^0 \rightarrow \pi^0 \mu^+ \mu^-$, 
$B^0 \rightarrow \eta e^+ e^-$, and $B^0 \rightarrow \eta \mu^+ \mu^-$.

%% file: introduction.tex
\section{Introduction\label{sec:introduction}}
In the  standard model (SM), the  decays \B\to\pill\ ($\pi=\pipm,\piz$
and  $\ell  = \electron,\mu$)  and  \Bz\to\etall  proceed through  the
quark-level  flavor-changing neutral  current (FCNC)  process \btodll.
Since all  FCNC processes are forbidden  at tree level in  the SM, the
lowest  order  diagrams representing  these  transitions must  involve
loops.  For  \btodll, these  are the electroweak  semileptonic penguin
diagrams (Fig.  1(a)) and the  $W^+W^-$ box diagrams (Fig.  1(b)).  The
\btodll transition is  similar to \btosll, but its  rate is suppressed
by  the  ratio  $|V_{td}/V_{ts}|^2\approx  0.04$  where  $V_{td}$  and
$V_{ts}$  are elements of  the Cabibbo-Kobayashi-Maskawa  quark mixing
matrix   \cite{Cabibbo:1963yz,   Kobayashi:1973fv}.    
The predicted branching fractions for the $\Bp\to\pipll$ decay modes lie in the
range of $(1.4\operatorname{--}3.3)\times 10^{-8}$, when the dilepton mass
regions near the \jpsi and \psitwos are excluded in order to remove decays that
proceed through the intermediate charmonium resonances.  The largest source of
uncertainty in these predictions arises from knowledge of the $B\to\pi$ form
factors \cite{Aliev:1998sk,    Wang:2007sp, Song:2008zzc}.   These branching  fractions imply  that  5--15 events
occur for each \B\to\pill\ decay  channel in the \babar\ data set (471
million $\BB$  pairs).  The predicted  $\Bz\to\etall$ branching fractions
lie in the range $(2.5\operatorname{--}3.7)\times 10^{-8}$ where again
the dominant uncertainty is due to lack of knowledge of the $\Bz\to\eta$ form
factors \cite{Erkol:2002nc}.  
\begin{figure*}
 \label{fig:SLPenDiagram}
  \begin{center}
   \centerline{\includegraphics[width=1.5\columnwidth]{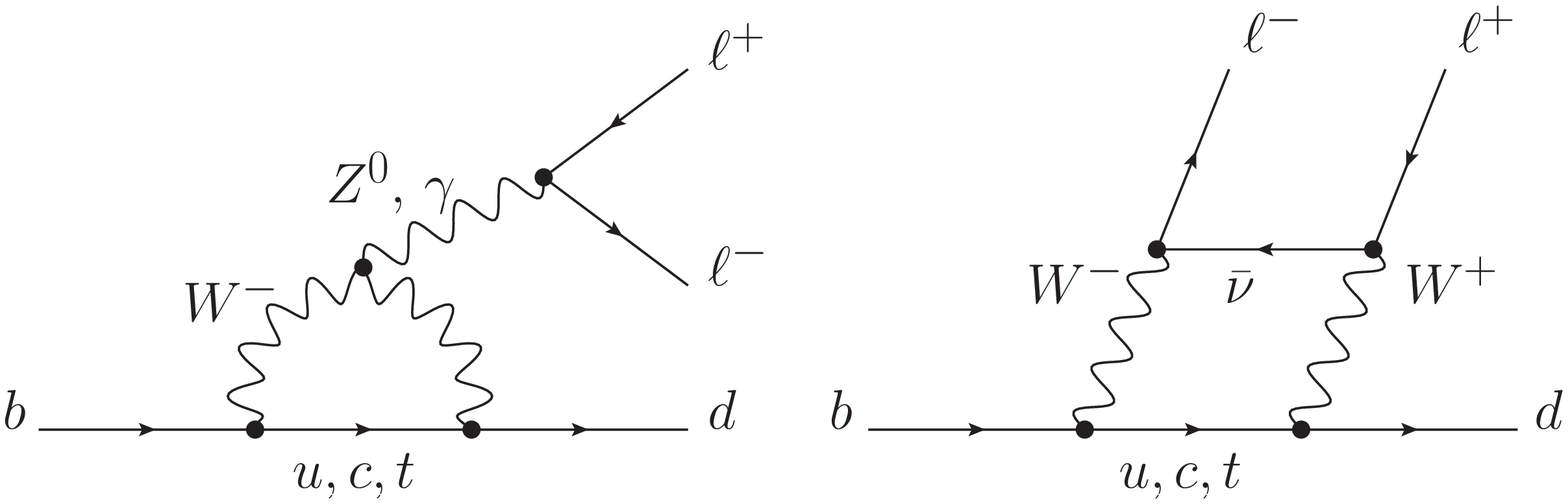}}
    \setlength{\unitlength}{1cm}
    \begin{picture}(0,0)
     \put(-3.80,0.025){{\bf (a)}}
      \put(2.75,0.025){{\bf (b)}}
   \end{picture}
    \caption{Lowest order Feynman  diagrams describing the quark level
      \btodll  transition   in  \B   meson  decay: (a)
      electroweak   penguin   diagrams and (b) $W^+W^-$   box   diagrams.}
  \end{center}
\end{figure*}

Many  extensions  of  the  SM  predict the  existence  of  new,  heavy
particles  which couple  to the  SM fermions  and bosons.   The
 \btodll  and \btosll decays  provide a promising  avenue in
which  to search  for  New  Physics (NP).   Amplitudes  from these  NP
contributions can interfere with  those from the SM, altering physical
observables  (\eg\  decay rates,  \CP,  isospin, and  forward-backward
asymmetries) from the  SM predictions \cite{Wang:2007sp, Aliev:1998sk,
  PhysRevD.61.074024,  Beneke:2001at,  Beneke:2005,  Feldmann:2002iw}.
Measurements in  the \pill\ and \etall systems  complement and provide
independent  searches  of  NP   from  those  in  the  $K^{(*)}\ellell$
channels
~\cite{PhysRevD.86.032012,PhysRevD.102.091803,PhysRevLett.103.171801,PhysRevLett.108.081807,JHEP.07.133,
hepex.1209.4284,hepex.1210.4492},  
as  physics  beyond  the  SM may  have  non-trivial  flavor
couplings  \cite{Davidson:1994}.   The measurement  of
observables as a function of the square of the invariant dilepton mass
$q^2=\mll^2$  for exclusive  \btodll decay  modes
allows for more  thorough tests of SM predictions  and deeper probes
for NP but is currently not possible due to the size of the data set.

Only one \btodll decay has  been observed to date, with LHCb measuring
the   \BtoPiMuMu   branching fraction to   be  $(2.4\pm   0.6\pm   0.2)\times
10^{-8}$~\cite{LHCb:2012}.    Both   \babar~\cite{Aubert:2007mm}   and
Belle~\cite{Wei:2008nv}      have      performed     searches      for
\B\to\pill\  decays, but  have  observed no  significant signal.   For  the
\pill\  modes, the smallest  upper limits from the \B factories lie  within an  order of
magnitude  of  the  SM  predictions  \cite{Aliev:1998sk,  Wang:2007sp,
  Song:2008zzc}  and  are beginning  to  exclude  portions  of the  NP
parameter  space.  No  previous  searches for  \Bz\to\etall have  been
reported.  Observation of \btodll decays at the \B\ factories is currently limited by the size of the
available data sets.  Additionally, for \Bp\to\pipll, background from
\Bp\to\Kll decays where the kaon is misidentified as a pion must be treated
carefully as \Kll can appear very signal-like and occurs at a rate approximately
25 times the expected \Bp\to\pipll rate.

In  this  article  we report  on  our  study  of the  \B\to\pill\  and
\Bz\to\etall  decays using the full \babar\ data set, presenting branching  fraction upper  limits for
the modes \BtoPiEE, \BtoPiZEE, \BtoPiMuMu, \BtoPiZMuMu, \BtoEtaEE, and
\BtoEtaMuMu.    Charge  conjugation   is  implied   throughout  unless
specified   otherwise.   We   also  present   upper  limits   for  the
lepton-flavor   averaged   modes   \Bp\to\pipll,   \Bz\to\pizll,   and
\Bz\to\etall,  where  we  constrain  the  \epem  and  \mumu  branching
fractions  to be equal;  for the isospin  averaged modes  $\B\to\piee$ and
$\B\to\pimm$, where the \Bp\to\pipll\ decay rate is constrained to
be  twice   the  \Bz\to\pizll\  decay   rate;  and  for the   isospin  and
lepton-flavor averaged  mode \B\to\pill.  For the lepton-flavor averaged
measurements, we neglect differences in available phase space due to the
difference between the electron and muon masses.  The  branching fractions are
based  on  signal  yields  that  are extracted  through  an  unbinned,
extended  maximum likelihood  fit  to two  kinematic variables.  All selection
criteria are determined before the fit was performed on data, \ie\ the
analysis is performed ``blind''.   

%% file: expsetup_dataset.tex
\section{\babar\ Detector, Simulation, and Data Sets\label{sec:expsetup_dataset}}
The   results  of   this  analysis   are  based   upon  a   sample  of
\epem\to\FourS\to\BB    interactions   provided    by    the   \pep2
asymmetric-energy storage rings and  collected by the \babar\ detector
located  at  SLAC  National  Accelerator  Laboratory.   The  \babar\  data  sample
corresponds to  an integrated  luminosity of 428\invfb  containing 471
million \BB decays.  This is the full data set collected at the \FourS
resonance. A detailed description
of  the  \babar\ detector  can  be  found elsewhere  \cite{Aubert:2001tu}.   Charged
particle momenta are measured  with a five-layer, double-sided silicon
vertex tracker  and a 40-layer drift chamber  operated in proportional
mode.  These two  tracking systems are immersed in  the 1.5 T magnetic
field  of  a   superconducting  solenoid.   A  ring-imaging  Cherenkov
detector with  fused silica radiators, aided by  ionization loss \dedx
measurements  from  the tracking  system,  provides identification  of
charged particles.  Electromagnetic showers from electrons and photons
are  detected with  an electromagnetic  calorimeter  (EMC) constructed
from  a finely  segmented array  of thallium-doped  CsI scintillating
crystals. The  steel flux return  of the solenoid (IFR)  was initially
instrumented  with  resistive  plate  chambers  (RPCs)  and  functions
primarily to  identify muons.  For the later data  taking periods, the
RPCs of the IFR were replaced with limited streamer tubes and brass to
increase absorption.

The \babar\  Monte Carlo (MC)  simulation utilizes the  \geant package
\cite{Agostinelli:2002hh}   for  detector   simulation,   and  \evtgen
\cite{Lange:2001uf} and {\tt Jetset7.4} \cite{Sjostrand:1995iq} for \BB and
\epem\to\qqbar  $(q=u,d,s,c)$ decays,  respectively.  The \BB and
continuum  (\epem\to\qqbar,  $q=u,d,s,c$)  MC samples correspond to  an  effective
luminosity of about ten times  the data sample collected at the \FourS
resonance.  Simulated  \B\to\pill\ signal decay  samples are generated
according  to the  form-factor model  of  Ref.~\cite{Ball:2003}, with  the
Wilson     coefficients    taken     from    Refs.~\cite{PhysRevD.66.034002,
  PhysRevD.65.074004,   Bobeth:2000},   and   the   decay   amplitudes
calculated in  Ref.~\cite{PhysRevD.61.074024}.  The \Bz\to\etall  signal MC
samples  utilize the  same  kinematics, Wilson  coefficients, and  form-factor 
model as  the \pill\ modes.  The effects of  the choice of form-factor model 
and the values  of the Wilson coefficients are considered
as  sources of systematic  uncertainty in  the signal  efficiency.  We
also make use of  simulated $\B\to K^{(*)}\ellell$, $\B\to\jpsi X$, and
$\B\to\psitwos X$ samples.  Signal efficiencies,  as well as parameters of the
fit  model, are  determined from  signal and  $K^{(*)}\ellell$  MC data
sets.  The $\B\to\jpsi X$ and $\B\to\psitwos X$ MC samples allow us to
study background  from these  decays and also  serve as the  data sets
from  which  we fix  the  parameters  of the  fit  model  used in  the
$\B\to\jpsi  \pi/\eta$  fit   validation, as described later.  

%% file: evtreco_candselec.tex
\section{Event Reconstruction and Candidate Selection\label{sec:evtreco_candselec}}  
Event reconstruction  begins by building dilepton  candidates from two
leptons (\epem or \mumu).  Leptons are selected as charged tracks with
momenta  in  the  laboratory   reference  (Lab)  frame  greater  than
300\mevc.  Loose particle  identification (PID) requirements are placed upon
the two  leptons.  More stringent PID requirements are applied later, and the optimization
of these selection criteria is discussed in Section~\ref{sec:nn_cut_opt}.  The lepton pair is fit
to a common  vertex to form a dilepton  candidate with the requirement
that  $\mll<5.0\gevcc$ \cite{Hulsbergen:2005}.   We  also place a loose
constraint on the $\chi^2$ probability of the vertex  fit by requiring 
it to be greater than $10^{-10}$.
For  electrons, we apply  an algorithm  which associates  photons with
electron candidates in an attempt to recover energy lost through bremsstrahlung,
allowing at most one photon  to be associated with each electron.  The
photon trajectory  is required  to lie within  a small cone of opening angle 0.035\rad about the
initial momentum vector of the  electron, and the photon energy in the
Lab  frame must  be greater  than 30\mev.  Additionally,  we suppress
background from photon conversions by requiring that the invariant mass of the electron
(or positron) paired with any other oppositely charged track in the
event be greater than 30\mevcc.

Charged  pion  candidates are  charged  tracks  passing pion PID
requirements which retain approximately 90-95\% of charged pions and only 2-5\%
of charged kaons.   We  reconstruct \piz  candidates from  two
photons with  invariant diphoton mass  $m_{\gaga}$ lying in  the range
$115 < m_{\gaga} < 150\mevcc$.  A minimum value of 50\mev is required for
the Lab energy  of each photon.  Photons are  detected as EMC clusters not
associated with a charged track. The clusters are also required to have a lateral  shower profile  consistent 
with originating from  a photon.   We reconstruct
$\eta$ as $\eta\to\gaga$  (\etagg) and $\eta\to\threepi$ (\etathreepi),
which constitute  39.3\% and 22.7\% of the  $\eta$ branching fraction,
respectively.  As in the case of the \piz, we require the \etagg\ photon daughters to
have energy greater  than 50\mev in the Lab  frame.  Additionally, the
photon              energy             asymmetry             $A_\gamma
=|E_{1,\gamma}-E_{2,\gamma}|/(E_{1,\gamma}+E_{2,\gamma})$ must be less
than 0.8, where $E_{1,\gamma}$  and $E_{2,\gamma}$ are the energies of
the photons in the Lab frame.  The invariant diphoton mass must lie in
the range  $500 < m_{\gaga}  < 575\mevcc$.  For \etathreepi,  the pion
candidates are fit  to a common vertex to form an $\eta$ candidate.  In 
the fit the $\eta$ candidate mass is constrained to the nominal $\eta$ 
mass, while the invariant three-pion mass is required
to lie in the range $535 < m_{3\pi} < 565\mevcc$.

\B candidates are reconstructed from  a hadron candidate (\pip, \piz, \etagg, or
\etathreepi) and a dilepton candidate (\epem or \mumu).  The hadron and dilepton candidates
are fit to  a common vertex, and the entire decay  chain is refit.  We
make use of two kinematic, Lorentz-invariant quantities, \mes and \DeltaE, defined as
\begin{eqnarray}
  \mes & = & \sqrt{(s/2 + \vec{p}_B\cdot\vec{p}_0)^2/E^2_0-p^2_B}\\
  \DeltaE & = & (2q_Bq_0-s)/2\sqrt{s}
\end{eqnarray}
where $\sqrt{s}=2E^{*}_{\rm beam}$ is the total energy of the \epem system in the
center of mass (CM) frame, $q_B$ and $q_0=(E_0,\vec{p}_0)$ are the four-vectors representing the
momentum of the \B candidate and of the \epem system, respectively, and $\vec{p}_B$ is
the three-momentum of the \B candidate.  In the CM frame, these expressions
simplify to
\begin{eqnarray}
  \mes    & = &\sqrt{E^{*2}_{\rm beam}-{p^{*}_{B}}^2}\\
  \DeltaE & = & E^{*}_B-E^{*}_{\rm beam}
\end{eqnarray}
where the asterisk indicates evaluation in the CM frame.
These variables make use of precisely measured beam quantities.  All \B candidates
are  required   to  have  $\mes>5.1\gevcc$  and  $-300   <  \DeltaE  <
250\mev$.   The distributions of these  two  variables  are  later  fit  to  extract  the
\pill\ and \etall branching fractions.

A large  background is present from $\B\to\jpsi  X$ and $\B\to\psitwos
X$ decays where \jpsi and \psitwos decay to \ellell.  Here $X$ represents a
hadronic state, typically $\pi$, $\eta$, $\rho$, or $K^{(*)}$.  These events are
removed   from  our   data  sample   by  rejecting   any   event  with a value of
\mll\ consistent with originating from a \jpsi or \psitwos decay.  The
rejected  \jpsi events  are useful  as they  provide a  control sample
which can be used to test the fit model.  We also use these samples to
estimate  systematic  uncertainties  and  to correct  for  differences
between data and MC selection efficiencies.  For the electron modes we
reject events in  the following regions about the  \jpsi mass: $2.90 <
\mll  < 3.20\gevcc$, or  $m_{ee} <  2.90\gevcc$  and $\DeltaE  < m_{ee}  c^2-
2.875\gev$.   For  the  muon  modes  the  region is  $3.00  <  m_{\mu\mu}  <
3.20\gevcc$, or  $m_{\mu\mu}  <  3.00\gevcc$  and  $\DeltaE <  1.11 m_{\mu\mu}  c^2  -
3.31\gev$.  The rejection  region about the \psitwos mass  is the same
for  electrons  and  muons: $3.60  <  \mll  <  3.75 \gevcc$, or $\mll  <
3.60\gevcc$ and $\DeltaE < \mll c^2 - 3.525\gev$.  Introducing \DeltaE
dependence into the region boundaries allows us to account for some of
the  effects   of  track  mismeasurement   and  energy  lost   due  to
bremsstrahlung.

The  largest source of  background comes  from random  combinations of
particles from continuum  events or semileptonic \B and  $D$ decays in
\BB events.  Continuum events tend to be jet-like as the $\qqbar$ pair
is produced  back-to-back with relatively  large momentum in  the CM frame.
The topology  of \BB  decays is  more isotropic as  the \B  mesons are
produced nearly at rest in the \FourS rest frame.  Semileptonic decays
are characterized by  the presence of a neutrino,  \eg\ missing energy
in the event and non-zero total transverse momentum of the event. Due to the
differences in these two background types we train separate artificial
neutral networks (NNs) to reject each of them.  By selecting inputs to the NNs
which are independent of the final state we are able to train only one
NN for  each lepton flavor.   We do not  train separate NNs  for \pip,
\piz, and $\eta$.  This increases the size
of the  training samples, improving  the performance of the  NNs.  We train four
NNs: one to reject \BB background in the\epem modes, one to reject \BB\
background in the \mumu\ modes, one to reject continuum background in the \epem
modes, and one to reject continuum background in the \mumu\ modes.  

The signal  training  samples  were   assembled  from  equal  portions  of
correctly      reconstructed      $\Bp\to\pipll$,      $\Bz\to\pizll$,
$\Bp\to\rhopll$,     $\Bz\to\rhozll$,     $\Bz\to\etagg\ellell$,    and
$\Bz\to\omega\ellell$  MC  events.  The size of the training samples, particularly
the background training samples, that could
be formed from events reconstructed as one of our signal modes was a limiting
factor in the performance of the NNs.  To increase the available statistics 
the other events from the $\rho\ellell$ and $\omega\ellell$ modes were added to
the training samples.  The $\rhop$ ($\rhoz$) was reconstructed as $\pi^+\pi^0$ 
($\pi^+\pi^-$) with two-pion invariant mass $m_{\pi\pi}$ in the range $0.455 <
m_{\pi\pi} < 1.095\gevcc$ ($0.475 < m_{\pi\pi} < 1.075\gevcc$).  The $\omega$
was reconstructed as $\pi^+\pi^-\pi^0$ and required to have a three-pion
invaraint mass lying within 50\mevcc of the nominal $\omega$ mass.  
No $\etathreepi\ellell$ events were used in the training due to very low
statistics for the background \BB and continuum samples for these modes.
For background we 
combined the MC data sets, either \BB or continuum depending upon the
classifier to be trained, from the six modes listed above and randomly
select events from this data set to form the training sample.  The performances of the NNs trained with
samples from  several different \btodll (global NNs)  modes were compared
with NNs  trained specifically for  each of our four  \B\to\pill\ modes
(single mode  NNs).  The background rejection  of the global  NNs at a
fixed signal efficiency was found to be similar to that of the single mode NNs.

The input variables  to the continuum NNs are  related mostly to event
topology and include  the ratios of Fox-Wolfram moments \cite{Fox:1978vw};
the cosine of the  polar angle of the thrust  axis \cite{Brandt:1964sa} of
the  event; the cosine of the polar angle of the thrust axis of the rest-of-the-event (ROE),  which  consists of  all
particles in  the event not  associated with the signal  \B candidate;
the momentum   weighted   polynomials   $L^j_i$ \cite{legendremom} computed using
tracks and EMC clusters;  the cosine of  the polar
angle of the \B candidate momentum; and the $\chi^2$ probability of the \B
candidate vertex fit.  The input  variables to the \BB NNs reflect the
effort to  reject background from  semileptonic \B and $D$  decays and
include $\mes$ and $\DeltaE$  constructed from the ROE; total momentum
of  the event transverse  to the  beam; missing  energy in  the event;
momentum of the ROE transverse to the beam direction; momentum of the ROE transverse to the thrust
axis  of the  event; cosine  of the  polar angle  of the  \B candidate
momentum; and  $\chi^2$ probability of  the \B candidate  and dilepton
candidate vertex fits.  The NN outputs show only weak correlation with the fit
variables \mes and \DeltaE.

Figure 2, as representative of the  several neural networks, shows  the output of the \epem \BB NN
for  a  sample  of  signal  and \BB background  \pipee\ MC events.
Also shown is the output of the \mumu continuum NN for a sample of signal and
continuum background \pizmm\ MC events.  
Requirements on  the NN  outputs are optimized  for each of  our eight
modes  to  produce  the  lowest  branching fraction  upper  limit.   A
description  of  the  optimization   procedure  is  given  in  Section
\ref{sec:nn_cut_opt}.

\begin{figure}
  \includegraphics[width=\columnwidth]{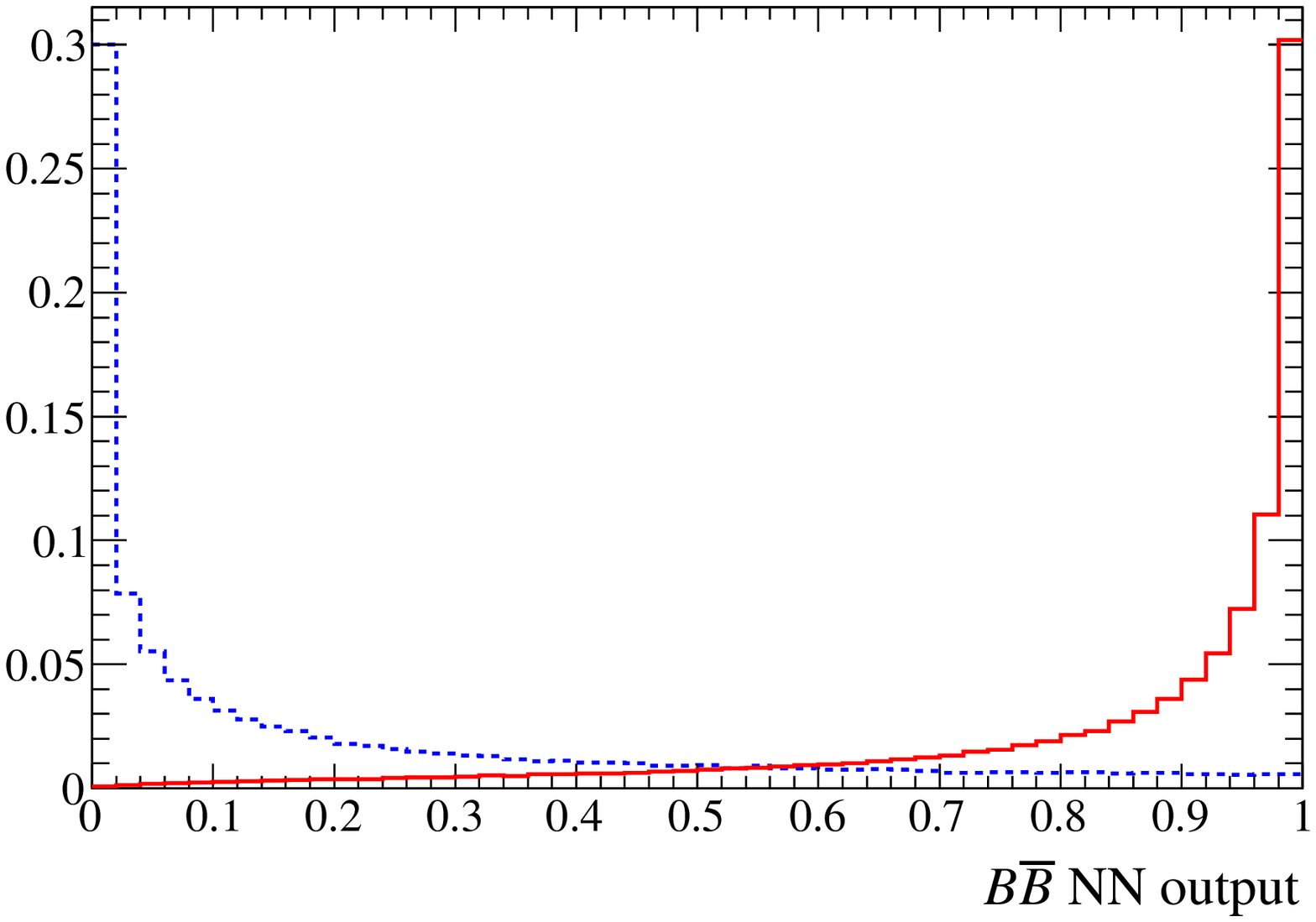}\\
  \includegraphics[width=\columnwidth]{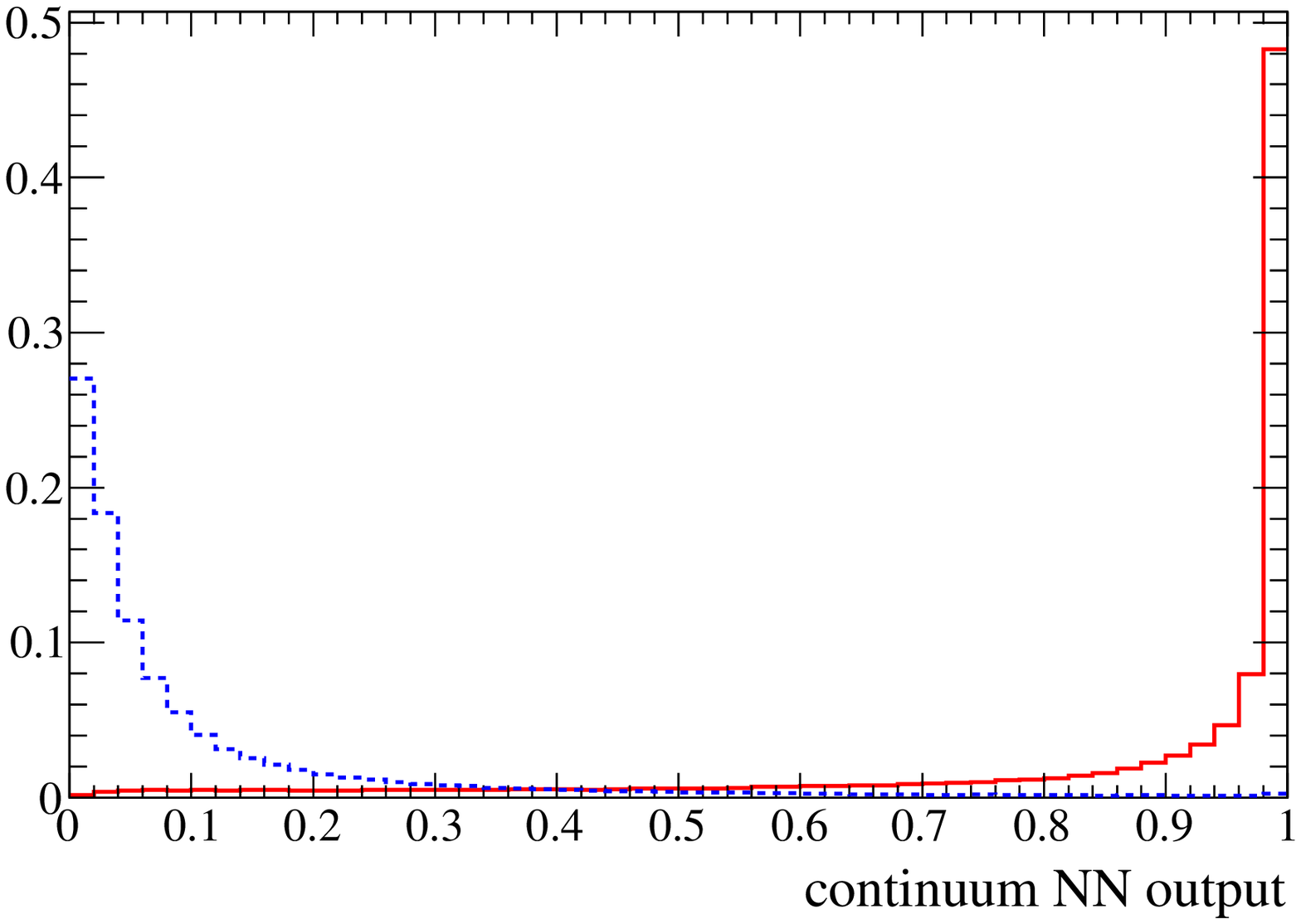}
  \setlength{\unitlength}{1in}
  \begin{picture}(0,0)
    \put(0.035,4.45){{\bf (a)}}
    \put(-1.65,3.85){\rotatebox{90}{{arbitrary units}}}
    \put(0.035,2.125){{\bf (b)}}
    \put(-1.65,1.525){\rotatebox{90}{{arbitrary units}}}
  \end{picture}
  \caption{Outputs of (a) the \epem \BB neural network for
    a sample of \BtoPiEE signal (solid red) and \BB background (dashed blue) MC events, and
    (b) the \mumu continuum neural network for a sample of \BtoPiZMuMu signal
    (solid red) and continuum background (dashed blue) MC events (color
    available online).  For both (a) and (b) the
    signal and background distributions are normalized to equal areas.}
\end{figure}

Due to their similarity to signal, $\B\to K^{(*)}\ellell$ decays constitute
a background that mimics signal by  peaking in either one or both \mes
and \DeltaE.  The \btosll transition occurs at a rate approximately 25
times  greater  than  the  SM   \btodll  rate,  and  due  to  particle
misidentification  and event  misreconstruction,  its contribution  is
expected to be of the same order as the \pill\ signal in the \babar\ data sample.  In the charged
pion  modes, $\Bp\to\Kll$  peaks  in  \mes as  \pipll\  signal but  in
\DeltaE near $-70\mev$  due to the misidentification of  the kaon as a
pion.  There are also contributions from $\Bz\to\KS\ellell$, where one of the pions
from the \KS decay is missed, and from $\B\to\Kstar(\to\Kp\pi)\ellell$, where the pion from
the \Kstar decay is missed.  In the
case  of   \KSll,  the  remaining   pion  and  the  two   leptons  are
reconstructed as  \pipll.  For $\Kstar(\to\Kp\pi)\ellell$,  the \Kp is
misidentified as \pip and reconstructed with
the two  leptons as \pipll.   In both cases,  the decays peak  in \mes
like signal but  at \DeltaE $<-140\mev$ due to  the missing pion.  For
\Kstarll the \DeltaE peak occurs at  even lower values due to the kaon
misidentification.  For $\Bz\to\pizll$,  there is a similar background
from  $\Bz\to\KS(\to\pi^0\pi^0)\ellell$  decays   where  one  \piz  is
reconstructed  along with  the lepton  pair as  \pizll.   These events
produce a peak in \mes in the same location as \pizll signal, but peak
at smaller values  of \DeltaE due to the missing  pion from the decay.
In  all three  cases  (\Kll in \pipll, $\KS(\to\pipi)\ellell$ and
$\Kstar(\to\Kp\pi)\ellell$ in \pipll,
and  $\KS(\to\pi^0\pi^0)\ellell$ in \pizll) we include  a separate  component in
the fit model to account for the corresponding contribution.

For  \piee\  and \etaggee,  there  is  an  additional background  that
originates   from   two-photon   events,   given   by   the   process
$\epem\to\epem\gaga\to(\epem)  \qqbar$ where $q$  is a \u, \d, or \s quark.
The background  is characterized by  a small transverse momentum  of the
pion  and  a  large  lepton-lepton opening  angle  $\theta_{\ellell}$.
There is also  a correlation between the polar  angles of the electron,
$\theta_{\en}$, and of the positron, $\theta_{\ep}$.  The \en  tends to be
in the  forward direction while  the \ep tends  to be in  the backward
direction, consistent  with the  \epem beam particles  scattering into
the detector.  Events  of this type are rejected using the following requirements.  For \pipee, \pizee, and \etaggee\ we
require   $p^{*}_{\rm  had}>750\mevc$   and   $N_{\rm  trk}>4$   where
$p^{*}_{\rm had}$ is the hadron  momentum in the CM frame and $N_{\rm trk}$
is  the number  of charged  tracks  in the  event.  Additionally,  for
\pipee\       we       require      $E_{1,{\rm       neut}}<1.75\gev$,
$\cos\theta_{\ellell}>-0.95$,        and        $\theta_{\en}        >
(0.57\,\theta_{\ep}-0.7\rad)$ where $E_{1,{\rm neut}}$  is the energy of the
highest  energy   neutral  cluster  in the event in  the   Lab  frame.   Similarly,
\pizee\      candidates     must      satisfy      $\theta_{\en}     >
(0.64\,\theta_{\ep}-0.8\rad)$, and \etaggee\ candidates are required to have
$\theta_{\en}          >          (0.6\,\theta_{\ep}-0.55\rad)$          and
$\cos\theta_{\ellell}>-0.95$.   These   criteria  were  determined  by
maximizing   the  quantity   $\varepsilon/\sqrt{N_{\rm   SB}}$,  where
$\varepsilon$ is the signal efficiency  and $N_{\rm SB}$ is the number
of events lying in the sideband  region $5.225 < \mes < 5.26\gevcc$ in
data. We assume that the two-photon background in the \mes\ sideband
occurs similarly to the two-photon background in the region $\mes > 5.26\gevcc$. The 
optimization  was  carried out  with  all other  selection
criteria applied, including those on the NN outputs.

To guard  against possible  background from $B\to  D\pi$ and  $\B\to D
\eta$ decays where  $D\to K \pi$, $\pi\pi$, or  $\eta\pi$ and the kaon
or pions are misidentified as leptons, we assign the lepton candidates
either a kaon or pion mass and discard any event with a combination of
$\mumu$, $\mu^{\pm}\pi$, or $\mu^{\pm}\eta$ with invariant mass in the
range (1.83--1.89)\gevcc.  The probability  of misidentifying a hadron as
an  electron is  negligible, and  this requirement  is  therefore only
applied to the \mumu modes.

Hadronic decays such as  $\Bp\to\pi^+\pi^-\pi^+$, where two pions are
misidentified as  muons, peak  in both \mes  and \DeltaE  similarly to
signal due  to the  relatively small difference  between the  pion and
muon  masses.   This  hadronic  peaking  background is  modeled  by  a
component  in the fit.   A dedicated  data control  sample is  used to
determine  its normalization  and shape.   This sample  is constructed
from events where one  lepton candidate passes the muon identification
requirements but the other does  not.  The events in these samples are
weighted with particle misidentification probabilities determined from
control samples in  \babar\ data.  Studies of MC samples
indicate that this background is consequential only for the \pimm\ modes.

After  applying all  selection criteria  there are  sometimes multiple
candidates within a given mode remaining in an event.  This occurs for
approximately 20--25\% (35--40\%) of  \pipee\ and \pizee\ (\etaggee\ and
\etathpiee)   candidates,  and   5--10\%  (25--30\%)   of   \pipmm\  and
\pizmm\ (\etaggmm\ and \etathpimm)  candidates.  There tend to be more
events containing  multiple candidates in  the \epem modes due  to the
bremsstrahlung  recovery.  For instance, there may be multiple candidates arising
from the same \pipee\ combination where the bremsstrahlung photons
associated with the \ep or \en are different.

To choose  the best  candidate we  construct a
ratio $\mathcal{L}_R$ from the  \BB and continuum NN classifier output
distributions  of  the  signal  and  background  samples.   The  ratio
$\mathcal{L}_R$ is defined as
\begin{equation}
  \mathcal{L}_R(x,y) = \frac{\mathcal{P}^{\rm sig}_{\BB}(x) + \mathcal{P}^{\rm
      sig}_{\rm cont}(y)}{(\mathcal{P}^{\rm sig}_{\BB}(x) + \mathcal{P}^{\rm
      sig}_{\rm cont}(y)) + (\mathcal{P}^{\rm bkg}_{\BB}(x) + \mathcal{P}^{\rm
      bkg}_{\rm cont}(y))}
\end{equation}
where  $\mathcal{P}^{\rm sig}_{\BB}(x)$ ($\mathcal{P}^{\rm sig}_{\rm cont}(y)$) is
the  probability that  a signal  candidate  has a  \BB (continuum)  NN
output      value     of      $x$      ($y$).      The      quantities
$\mathcal{P}^{\rm bkg}_{\BB}(x)$   and  $\mathcal{P}^{\rm bkg}_{\rm cont}(y)$  are
defined  analogously for  background  events.  Signal-like  candidates
have  values  of $\mathcal{L}_R$  near  1  while more  background-like
candidates have values near 0.   If multiple candidates are present in
an  event,  we  choose  the  candidate  with  the  greatest  value  of
$\mathcal{L}_R$  as the  best candidate.   For events containing multiple
candidates, this procedure  chooses the
correct candidate  approximately 90--95\%  of the time for \pill\ and 75--80\% of
the time for \etall.  The ratio $\mathcal{L}_R$ is used only to select a best candidate.

%% file: fit_model.tex
\section{Branching Fraction Measurement and Upper Limit Calculation\label{sec:fit_model}}
Branching fractions are extracted through an unbinned extended maximum
likelihood  fit to \mes  and \DeltaE  with the  fit region  defined as
$\mes > 5.225 \gevcc$ and $-300 < \DeltaE < 250 \mev$.  The probability
density functions  (PDFs) in the fit model  contain several components
corresponding  to the  different contributions  in the  data  set.  To
model the various components,  we use a combination of products of
one-dimensional  parametric   PDFs,  two-dimensional  histograms,  and
two-dimensional non-parametric shapes  determined by a Gaussian kernel
density estimation algorithm (KEYS PDF) \cite{Cranmer:2000}.  For components
that are described by the product of one-dimensional PDFs, we are allowed to
use such a model because \mes and \DeltaE are uncorrelated for these components.

\subsection{\Bp\to\pipll\label{sec:pipll_fit}}
The  \pipll\   fit  model  involves  four   components:  signal,  \Kll
background,  $\KS/\Kstarll$ background,  and  combinatoric background.
There  is  an  additional  component in  \BtoPiMuMu  representing  the $\Bp\to\pi^+\pi^+\pi^-$ 
hadronic peaking  background.  The \Kll background  arises from decays
where the kaon is misidentified  as a pion.  The \Kp misidentification
rate is such that the  \Kll background in \pipll\ is approximately the
same  size  as  the  expected   SM  \pipll\  signal.   Since  the  \Kp
misidentification  probability is  well  measured, it  is possible  to
measure this background contribution  directly from our data.  This is
done  by simultaneously  fitting two  data samples,  comprised  by the
\Bp\to\pipll\ candidates and the \Bp\to\Kll candidates in our data set.
The \Kp misidentification background  to \Bp\to\Kll is included in the
fit  at  a  level  fixed  to  the \Bp\to\Kll  yield  using  the  known
misidentification probability  (which depends  on the momentum  of the
kaon).  The  \Bp\to\Kll branching fraction  that is measured  from the
simultaneous  fit of  the  \Bp\to\pipll\ and  \Bp\to\Kll data  samples
provides  an  additional  validation  of  our  procedure,  since  this
branching fraction has been previously measured \cite{Nakamura:2010zzi}.

The  \Kll  sample  is  selected   in  exactly  the  same  way  as  the
\pipll\ sample except the charged pion identification requirements are
reversed  and the  \jpsi and  \psitwos rejection  window  includes the
following  regions:  $m_{ee}>3.20\gevcc$  and  $1.11 m_{ee} c^2  -  3.67  <
\DeltaE  <m_{ee} c^2  - 2.875\gev$  for  \pipee\ surrounding  the \jpsi  mass,
$m_{\mu\mu}>3.20\gevcc$  and $1.11 m_{\mu\mu}  c^2 -  3.614 <  \DeltaE <m_{\mu\mu}  c^2 -
2.925\gev$ for \pipmm\ surrounding the \jpsi mass, and $\mll>3.75\gevcc$ and
$1.11\mll c^2 - 4.305 < \DeltaE  < \mll c^2 -3.525\gev$ for both modes
surrounding the \psitwos mass.  Also,  the \DeltaE window is $-200 < \DeltaE
< 250\mev$ for \Kee\ and $-100 < \DeltaE < 250\mev$ for \Kmm.

The \pipll\ and \Kll background  \mes and \DeltaE distributions are modeled by 
products of one-dimensional PDFs.  The \pipll\ signal and \Kll background \mes distributions are described
by a  Crystal Ball  function \cite{Oreglia:1980}.  The  \pipee\
\DeltaE signal distribution is modeled by  the sum of  a Crystal Ball
function  and  a  Gaussian  which  share  a  common  mean,  while  the
\pipmm\ signal  and both the  \Kee and \Kmm \DeltaE  distributions are
modeled by  a modified Gaussian with tail  parameters whose functional
form is given by
\begin{equation}
  f(\DeltaE) = \exp{\left[ -\frac{(\DeltaE-\mu)^2}{2\sigma_{L,R}\alpha_{L,R}+\alpha_{L,R}(\DeltaE-\mu)}\right]}
  \label{eqn:Cruijff}
\end{equation}
where $\sigma_{L}$ and $\alpha_{L}$ ($\sigma_{R}$ and $\alpha_{R}$) are 
the width and tail parameters used when $\DeltaE<\mu$ ($\DeltaE>\mu$), respectively.
A two-dimensional  histogram models the contribution from
$\B\to\KS/\Kstar\ellell$ decays.  Combinatoric background is described
by the  product of  an ARGUS function  \cite{ARGUS:1990} in  \mes with
endpoint fixed to 5.29\gevcc and a second-order polynomial in \DeltaE.
The  \pipmm\ hadronic  peaking background  component is  modeled  by a
two-dimensional KEYS PDF \cite{Cranmer:2000}.

The PDF fit  to the \Kll sample contains a  similar set of components.
Signal \Kll distributions are modeled by the product of a Crystal Ball
function in  \mes and the  line shape of Eq. 6 in
\DeltaE.  The  contribution from other \btosll decays  is dominated by
$B\to\Kstar(K^+\pi)\ellell$  where  the  pion   is  lost.   We  use  a
two-dimensional  histogram  to  model this  background.   Combinatoric
background is modeled by the product of an ARGUS distribution in \mes,
and by an exponential function for \Kee and a second-order polynomial for \Kmm
in \DeltaE.  A KEYS PDF models the hadronic peaking background in \Kmm.

In  both  the \pipll\  and  \Kll  PDFs,  the signal  and  combinatoric
background  yields float  along with  the shapes  of  the combinatoric
background PDFs.  The  \Kll background yield in the  \pipll\ sample is
constrained so  that the $\Bp\to\Kll$ branching  fractions measured in
the \pipll\ and  \Kll samples are equal.  All  fixed shapes and yields
are  determined from  exclusive  MC samples  except  for the  hadronic
peaking background  which uses a data  control sample.  Normalizations
of the  $\KS/\Kstarll$ component  of the \pipll\  PDF and  of \Kstarll
component in the \Kll PDF  are fixed from efficiencies determined from
MC  samples   and  world   average  branching  fractions \cite{Nakamura:2010zzi}.

\subsection{\Bz\to\pizll\label{sec:pizll_fit}}
The \Bz\to\pizll\ signal  distribution is modeled by the  product of a
Crystal Ball function  in \mes and by the  line shape given in Eq.
\ref{eqn:Cruijff}       in       \DeltaE.        Background       from
$\Bz\to\KS(\to\pi^0\pi^0)\ellell$    decays    is    modeled   by    a
two-dimensional histogram.  The product of an ARGUS shape in \mes with
an   exponential function in  \DeltaE   models  the   combinatoric  background
distribution.   As  in  the  \pipmm\  and  \Kmm\  PDFs,  there  is  an
additional  component in  the \pizmm\  fit model  devoted  to hadronic
peaking background which is described by a KEYS PDF.

In the fit, only the signal \pizll\ and combinatoric background yields
along with the  ARGUS slope parameter and argument  of the exponential
float.    The  signal   and  $\KS(\to\pi^0\pi^0)\ellell$   shapes  are
determined     from     fits     to     MC    samples,     and     the
$\KS(\to\pi^0\pi^0)\ellell$  normalization   comes  from  efficiencies
taken  from MC  samples  and world  average  branching fractions \cite{Nakamura:2010zzi}.  The  shape and normalization  of the peaking
hadronic component are determined from a data control sample.

\subsection{\Bz\to\etall\label{sec:etall_fit}}
The  \etall\ fit model  is simple,  consisting of  only three
components,   and  is  the   same  for   all  four   \etall\  channels.  
The  signal component is  modeled by the product  of a
Crystal  Ball function  in \mes  and the  line shape of  Eq.  4 in
\DeltaE.  We include a component  for events containing a signal decay
where the signal \B is incorrectly reconstructed, which we refer to as
self-cross-feed.  In these events the signal decay is typically reconstructed as
a combination of particles from the \B decaying to our signal mode and the other
\B.  In most self-cross-feed events the dilepton pair is correctly reconstructed
and the hadron is misreconstructed.  The self-cross-feed contribution is represented by a two-dimensional
histogram  and its  normalization is  a fixed  fraction of  the signal
yield   with   the   fraction   determined  from   signal   MC.    The
self-cross-feed-to-signal   ratio   varies   from  0.1--0.15   for   the
\etagg\   channels  to   0.25--0.3  for   the   \etathreepi\  channels.
Combinatoric  background  is described  by  the  product  of an  ARGUS
function in \mes  and an exponential function in \DeltaE.   From studies of MC samples, we find no indication
of potential peaking background contributions from \btosll decays or any  other sources.  The
\etagg\ellell yield  and the \etathreepi\ellell  yield are constrained
in  the fit  to be  consistent with  the same  \Bz\to\etall\ branching
fraction.   The  signal yield,  combinatoric  background yield,  ARGUS
slope and exponential argument float in the fit.  All other parameters
are fixed from MC samples.

\subsection{Lepton-flavor averaged and isospin averaged fits\label{sec:averaged_fit}}
In addition  to branching fraction measurements and  upper limits for
the   $\B\to\pill$   and   $\Bz\to\etall$   modes  we   also   present
lepton-flavor  averaged,  isospin   averaged,  and  lepton-flavor  and
isospin averaged  results.  The lepton-flavor  averaged measurement of
$\BR(\Bp\to\pipll)$  is   the  branching  fraction   obtained  from  a
simultaneous fit  to the  \pipee\ and \pipmm\  samples subject  to the
constraint   $\BR(\BtoPiEE)=\BR(\BtoPiMuMu)$.    Here we have neglected the
difference between the electron and muon masses.  The  measurements   of
$\BR(\Bz\to\pizll)$ and $\BR(\Bz\to\etall)$ are 
subject to a similar set of constraints and are determined in an analogous
way.  The isospin averaged branching fraction $\BR(\B\to\piee)$ is the
measured  value of  $\BR(\BtoPiEE)$ after  simultaneously  fitting the
\pipee\ and \pizee\ samples subject to the constraint $\BR(\BtoPiEE) =
(\tau_{\Bz}/2\tau_{\Bp})\BR(\BtoPiZEE)$    where    $\tau_{\Bz}$   and
$\tau_{\Bp}$  are the mean  lifetimes of  the  neutral and
charged \B mesons, respectively \cite{Nakamura:2010zzi}.  An analogous
expression  is  applied for  the  $\BR(\B\to\pimm)$ measurement.   The
lepton-flavor and isospin averaged measurement of $\BR(\B\to\pill)$ is
the value of $\BR(\Bp\to\pipll)$ determined from a simultaneous fit to
all  four  samples subject  to  both  the  lepton flavor  and  isospin
constraints listed above.

\subsection{Upper limit calculation\label{sec:upper_limit}}
We  set   upper  limits  on   the  branching  fractions   following  a
method     which     utilizes     the     profile     likelihood.  
Upper limits at the $\alpha$ confidence level (CL) are set by scanning the
profile likelihood $\lambda$ as a function of the signal branching 
fraction to determine where $-2\ln\lambda$ changes by $\alpha$ percentile
of a $\chi^2$ random variable with one degree of freedom.  For $\alpha=0.9$
we look for a change in $-2\ln\lambda$ of 1.642.  If the measured branching fraction
is  negative, we begin  our scan  from zero  rather than  the minimum~\cite{Rolke:2004mj}.
This  is  a  conservative  approach  that  always  produces  physical,
\ie\ non-negative, upper  limits, even in the case  of low statistics.
Systematic uncertainties are incorporated into the limit by convolving
the profile  likelihood with  a Gaussian distribution whose  width is 
equal  to the total systematic uncertainty.

%% file: nn_cut_opt.tex
\section{Optimization of Selection\label{sec:nn_cut_opt}}
We  simultaneously optimize  the  selection criteria  for  the two  NN
outputs  and the  PID selection  criteria  for the  charged pions  and
leptons.   \babar\ employs algorithms  which use  outputs from  one or
more multivariate classifiers to identify charged particle species.  A
few (3-6) standard  selections on the outputs of  these algorithms are
used  to  identify  particles   of  a  given  species  with  different
efficiencies.   Greater  identification  efficiencies typically  imply
greater  misidentification rates.  Due  to this  trade-off, it  is not
clear  {\it  a  priori}  which  selection is  best  for  a  particular
analysis.  Therefore for each  charged particle type (\en, \mun, \pip)
we optimize the PID requirements  for the leptons and pions along with
the NN output criteria.

For the optimization we assume that    \B\to\pill\   and
\Bz\to\etall\ occur  near the center of the  branching fraction ranges
expected in the  SM.  Under this assumption, no statistically significant signal
is expected, and the selection is optimized to produce the  smallest branching fraction
upper    limit.  We divide the \BB and  continuum NN output space
into a  grid and generate 2,500 parametrically simulated  data sets per grid
point according to
our  fit model.   Each simulated  data set  is fit,  and a branching fraction upper limit is calculated.  The figure
of merit (FOM) for each  point is the average branching fraction upper
limit  determined  from the  2,500  data  sets, and we take the combination of PID  and NN output
selection  producing  the  smallest FOM as 
optimal.   

The  results of  the optimization  show that the upper  limits are  rather
insensitive  to the PID  selection.  Also,  in the  two-dimensional NN
output space, there is a  region about the optimal selection where the
FOM changes slowly, giving  confidence that our optimization procedure
is robust because the expected  limits do not depend critically on the
NN selection requirements.

The \epem (\mumu) modes use  the same electron (muon) selection, while
more  efficient charged  pion  selection is  favored  for \pipee\  and
\etathpiee\ than \pipmm\ and \etathpimm.  Tighter selection is favored
on the continuum  NN output than the \BB  NN output.  The optimization
favors looser  requirements for  the \etall modes  as the size  of the
background in these channels is much smaller than for \pill.

%% file: fit_validation.tex
\section{Fit Validation\label{sec:fit_validation}}
We  validate our  fit  methodology  in three  ways:  (1) generating  an
ensemble of  data sets from  our fit model  and fitting them  with the
same model (``pure pseudo-experiments''), (2) generating and fitting an
ensemble  of  data  sets  with  signal  events  from  the  \babar\  MC
simulation embedded into  the data set (``embedded   pseudo-experiments''),   (3)   extracting
$\B\to\jpsi\pi$  and $\Bz\to\jpsi\eta$  branching  fractions from  the
\babar\ data sample.

From our studies of both pure and embedded pseudo-experiments, we find
no significant source of bias  in our fit.  Distributions of branching
fractions and  their errors obtained  from fits to these  data sets are
consistent with expectations.

Measuring the $\Bp\to\jpsi\pip$, $\Bz\to\jpsi\piz$, $\Bz\to\jpsi\eta$,
and  $\Bp\to\jpsi\Kp$ branching  fractions  in the  control sample  of
vetoed charmonium events allows  us to validate our fit methodology
on  data.  We employ  the same  fit model  to extract  these branching
fractions as we do for  the \pipll, \pizll, \Kll, and \etall branching
fractions.  Fixed  shape parameters and yields  are determined through
fits to  exclusive MC samples.  We  find that all  measurements are in
good agreement with world averages~\cite{Nakamura:2010zzi}.

%% file: systematics.tex
\section{Systematic Uncertainties\label{sec:systematics}}
Systematic uncertainties 
are included in the branching fraction upper  limit calculation by
convolving the profile likelihood with a Gaussian whose width is equal
to the total systematic uncertainty.
The systematic uncertainties are divided into ``multiplicative'' uncertainties,
which scale with the true value of the branching fraction, and 
``additive'' uncertainties, which are added to the true value of the 
branching fraction, independent of its value.

\subsection{Multiplicative uncertainties}
We list the sources of multiplicative systematic uncertainty below and
their assigned values for each  of the \pill\ and \etall\ signal modes
in Table I.

\begin{table*}[!ht]
  \begin{center}
    \setlength{\extrarowheight}{1.8pt}
    \caption{Multiplicative systematic uncertainties  for the \pill\ and
      \etall\ modes.  The  lepton and \pipm PID and  NN output selection
      efficiency  correction uncertainties  are determined  using $\jpsi
      K^{(*)}$    control    samples,    while    the    tracking    and
      \piz/\etagg\ efficiency  correction and \B  counting uncertainties
      are taken  from dedicated \babar\ studies.   The total uncertainty
      is the sum in quadrature of the individual uncertainties.}
    \begin{tabular}{l c c c c c c c c}\hline\hline
                    & \pipee & \pizee & \pipmm & \pizmm & \etaggee & \etathpiee & \etaggmm & \etathpimm\\ \hline
      $N_{\BB}$     & 0.6\%  & 0.6\%  & 0.6\%  & 0.6\% & 0.6\% & 0.6\% & 0.6\% & 0.6\%\\ 
      \piz/\etagg\ eff.     & -      & 3.0\%  &  -     & 3.0\% & 3.0\% & 3.0\% & 3.0\% & 3.0\%\\
      Tracking eff. & 0.9\%  & 0.6\%  & 0.9\%  & 0.6\% & 0.6\% & 1.2\% & 0.6\% & 1.2\%\\ 
      lepton PID    & 1.3\%  & 1.3\%  & 1.4\%  & 1.4\% & 1.4\% & 1.4\% & 1.5\% & 1.5\%\\
      \pipm PID     & 2.5\%  & -      & 3.5\%  &  -    & -     & 2.3\% & -     & 3.7\%\\
      NN cut        & 1.4\%  & 1.3\%  & 1.5\%  & 1.5\% & 1.4\% & 1.3\% & 1.5\% & 1.5\%\\ 
      Wilson coeff. & 2.7\%  & 2.3\%  & 1.0\%  & 1.9\% & 3.1\% & 3.1\% & 0.3\% & 0.9\%\\
      FF model      & 9.1\%  & 7.7\%  & 0.7\%  & 7.1\% & 3.4\% & 1.3\% & 0.2\% & 1.6\%\\ \hline
      Total         & 10.1\% & 8.8\%  & 4.4\%  & 8.2\% & 5.9\% & 5.6\% & 3.8\% & 5.7\%\\ \hline\hline
    \end{tabular}
  \end{center}
  \label{tab:pillMultSyst}
\end{table*}

The  systematic uncertainty  in the  measured number  of \BB  pairs is
estimated to be 0.6\% \cite{McGregor:2008ek}.

The difference between the  \piz reconstruction efficiency in data and
MC has been  studied in \tautau decays where one  \mtau decays via the
channel  $\tau^\pm\to\epm\nunub$  and   the  other  \mtau  decays  via
$\tau^\pm\to\pipm\nu$  or   $\tau^\pm\to\rho^\pm\nu$  with  $\rho^\pm$
reconstructed as $\pipm\piz$.  The $\tau^\pm\to\rho^\pm\nu$ yields are
roughly   proportional  to  the   product  of   the  \pipm   and  \piz
reconstruction efficiencies,  while the $\tau^\pm\to\pi^\pm\nu$ yields
are  proportional   to  the  \pipm   reconstruction  efficiencies.   A
correction  proportional  to  the  ratio of  the  $\tau\to\rho\nu$  to
$\tau\to\pi\nu$ yields in the data and MC samples is applied  to better reproduce
the data reconstruction efficiency in MC simulation.  The uncertainty due to this
correction is estimated as 3.0\% per \piz.  We take the uncertainty in
the \etagg\ reconstruction  efficiency associated with this correction
to also be 3.0\% per \etagg.

A  correction to  the MC  tracking efficiency  was developed  from the
study of \tautau decays where  one \mtau has a single charged daughter
(1-prong  decays) allowing  the event  to be  identified as  a \tautau
event  and  the  other  \mtau  has three  charged  daughters  (3-prong
decays). By  measuring the  event yields where  the 3-prong  \mtau has
either  two or  three tracks  reconstructed, the  track reconstruction
efficiency can  be measured.  This  efficiency can be used  to correct
the  MC to  match the  efficiency  measured in  data.  The  systematic
uncertainty associated  with this correction is estimated  to be 0.3\%
per charged track  taken to be 100\% correlated  among tracks in the
event.

We  correct  for  the  difference  between the  lepton  PID  selection
efficiencies in data and MC  by measuring the $\Bp\to\jpsi\Kp$ yields in
data  and  $\jpsi\Kp$  MC  control  samples with  and  without  the  PID
selection  requirements applied  to both  leptons. The  ratios  of the
yields  are  used  to   correct  the  lepton  particle  identification
selection efficiency  derived from MC to match  data.  The
error  on  the  correction  is  taken  as  the  associated  systematic
uncertainty, which ranges from 1.3--1.5\%.  The available statistics in the samples used to 
calculate the correction determine the size of the error which is associated
with it.

In  an analogous  procedure,  we correct  for the difference between  the
charged pion  PID selection efficiency obtained by measuring signal  yields in
high statistics $\Bz\to\jpsi\Kstarz(\to  K^-\pi^+)$ data and exclusive
MC  control  samples with  and  without  pion  PID selection  criteria
applied.  A  correction is  derived and the  error  on the
correction is  taken as the associated  systematic uncertainty.  These
uncertainties are  approximately 2.5\% and  3.5\% for \epem  and \mumu
modes, respectively.

The high statistics of the  $\Bp\to\jpsi\Kp$ data and MC control samples
are again  exploited to  derive a correction  for the NN  output selection
efficiency on MC.  The $\jpsi\Kp$  signal yields were measured with only
the  \BB NN  output selection applied,  only the  continuum NN  output selection
applied, and with both selections applied.  The error on the correction
is  taken as  the associated  systematic uncertainty  and  ranges from
1.3--1.5\%.

We conservatively vary the Wilson coefficients  $C_7$, $C_9$, and $C_{10}$ from their
nominal  values of $-0.313$, $4.344$, and $-4.669$, respectively, by  a factor
of $\pm  2$ (\eg\ $C_7$ is varied to $-0.157$ and $-0.616$) and  generate  new simulated
samples  with   all  possible   combinations  of  the   varied  Wilson
coefficients.   For  each  varied  sample  we  apply  the  full  event
selection  and  calculate  the  efficiency  for  that  set  of  Wilson
coefficients,  taking  the  largest  relative difference  between  the
varied  Wilson  coefficient efficiencies  and  the  efficiency of  our
default model as the associated systematic uncertainty.

Simulated MC samples using several different form-factor models were
generated.   Ultimately, we  compare the  efficiency from  our default
model with the efficiency calculated  from the ``Set\ 2'' and ``Set\ 4''
form-factor   models  of  Ref.~\cite{PhysRevD.71.014015}.    The  maximum relative
difference  between our  default model  efficiency and  the efficiency obtained
with the
``Set\ 2'' and ``Set\ 4''form-factor models is  taken as
the associated systematic uncertainty.  There is a large variation in this
uncertainty from one mode to another.  The source of this effect is due in part
to the correlation between the NN output and \qsq.  Selection on the NN outputs
is mode dependent and therefore changes the \qsq dependence of the efficiency.
Also, for the modes \pipee, \pizee, and \etaggee\ we require that the hadron
momentum be greater than 750 \mevc in the CM frame.  The hadron momentum is
highly correlated with \qsq.  Removing events with small hadron momentum
also removes events with large \qsq.  Differences at large \qsq between the differential branching
fractions calculated using the default and alternative models lead to greater
sensitivity to the choice of form-factor model, and therefore larger
uncertainties associated with the choice of model.

The uncertainty  in the efficiency due to  the size of the
simulated MC samples is less than 0.1\% and is negligible.

\subsection{Additive uncertainties}
\begin{table*}[!ht]
   \setlength{\extrarowheight}{3.5pt}
    \centering
    \caption{Additive systematic  uncertainties for the  \B\to\pill\ and
      \Bz\to\etall\  channels. The total
      uncertainty   is  the   sum  in   quadrature  of   the  individual
      uncertainties.  All uncertainties are given in units of $10^{-8}$.}
    \begin{tabular}{l c c c c c c}\hline\hline
              Mode                                & \pipee & \pizee & \pipmm  & \pizmm   & \etaee            & \etamm   \\ \hline
      Fixed parameters                & $1.9$  & $0.2$  & $0.6$   & $0.3$  & 0.6 & 0.4 \\
      Non-parametric shapes           & $<0.1$ & $<0.1$ & $0.7$   & $0.5$  & 0.1 & 0.1 \\
      Hadronic peaking bkg yields     &  -     &  -     & $<0.1$  & $<0.1$ & -   & -   \\
      Non-hadronic peaking bkg yields &  -     &  -     & $<0.1$  & $<0.1$ & -   & -   \\ \hline
      Total                           & $1.9$  & $0.2$  & $0.9$   & $0.6$  & 0.6 & 0.4 \\ \hline\hline      
    \end{tabular}
 \label{tab:AddSyst}
\end{table*}

We consider  the following sources of  additive systematic uncertainty
with their values given in Table II.

The fixed parameters  of the \pill, \etall, and  \Kll signal and the \Kll
background  PDFs are  varied individually  within the  errors obtained
from  fits to exclusive  MC samples,  and the data  sample is
re-fit.  For  simultaneous fits we additionally  vary the efficiencies
within   their   uncertainties,   and   for  \etall\   we   vary   the
self-crossfeed-to-signal ratio by  $\pm 10\%$.  The size of the variation is 
arbitrary but conservative enough since the number of expected self-crossfeed 
events is at most 0.15.  The difference between
the branching fraction from this fit  and that from the nominal fit
is taken as the associated systematic uncertainty. We take the largest
change  in  the  branching  fraction  as  the  systematic  uncertainty
associated   with  each  fixed   quantity.   The   uncertainties  from
individual variations are summed in quadrature.

Non-parametric  PDFs include the  two-dimensional histograms  and KEYS
shapes.  We  vary the binning of the  two-dimensional PDFs, increasing
and decreasing  them by a factor  of two.  The data  sample is re-fit,
and  we take  the  largest change  in  the branching  fraction as  the
associated systematic  uncertainty.  For the KEYS  shapes, we increase
and decrease  the width  of the Gaussian  kernel used to  generate the
shapes and  take the largest change  in the branching  fraction as the
associated   systematic   uncertainty.     If   there   are   multiple
non-parametric PDFs, we add their associated uncertainties in quadrature.

The  hadronic peaking  background yields  are fixed  from  the control
sample  of hadronic  decays and  are varied  within  their statistical
uncertainties.  The  data sample  is re-fit, and  we take  the largest
change  in  the  branching   fraction  as  the  associated  systematic
uncertainty.

We fix  the $\KS/\Kstarll$  yield in the  \pipll\ PDF,  the $\Kstarll$
yield  in the  \Kll PDF,  and the  $\KSll$ yield  in the  \pizll\ PDF.
These values are determined  from efficiencies taken from exclusive MC
samples and the current world average branching fractions for these modes \cite{Nakamura:2010zzi}.  We vary
the yields  according to the  errors on their branching  fractions and
re-fit the data sample.  The change in the branching fraction from its
nominal value  is taken as the associated  systematic uncertainty.  In
Table II, these uncertainties are classified as ``Non-hadronic peaking
bkg yields''.

%% file: results.tex
\section{Results\label{sec:results}}
\begin{table*}[!ht]
  \begin{center}
    \setlength{\extrarowheight}{3.5pt}
    \caption{  \Btopill\   and  \Bz\to\etall  efficiencies, yields, branching   fractions,  and
      branching fraction upper limits at the 90\% CL.  The error on
      the yield is statistical.  The
      first error  quoted on the branching fractions  is statistical while
      the  second  is  systematic.   Branching  fraction  upper  limits
      include systematic uncertainties.}
    \begin{tabular}{ l c c c c }\hline\hline
      \multc{1}{c}{Mode} & \hspace{2em}$\varepsilon$\hspace{2em} & \hspace{3em}Yield\hspace{3em} &  \hspace{0.9cm} $\BR\,(10^{-8})$ \hspace{0.7cm} & Upper Limit $(10^{-8})$ \\\hline
      \BtoPiEE      & 0.199 & $4.2^{+5.7}_{-4.6}$ & $4.3^{+5.9}_{-4.7}\pm 2.0$ & 12.5\\
      \BtoPiZEE     & 0.163 & $1.0^{+3.2}_{-1.1}$ & $1.2^{+5.4}_{-4.0}\pm 0.2$ & 8.4\\
      \BtoEtaEE     &  &  & $-4.0^{+10.0}_{-8.0}\pm 0.6$ & 10.8 \\
      \hspace{1em}\BtoEtaGGEE   & 0.164 & $-1.2^{+3.1}_{-2.4}$ & & \\
      \hspace{1em}\BtoEtaThPiEE & 0.115 & $-0.5^{+1.2}_{-1.0}$ & & \vspace{0.2cm}\\
      \BtoPiMuMu    & 0.140 & $-0.5^{+3.1}_{-2.3}$ & $-0.6^{+4.4}_{-3.2}\pm 0.9$ & 5.5\\
      \BtoPiZMuMu   & 0.115 & $-0.2^{+2.0}_{-0.7}$ & $-1.0^{+5.0}_{-3.4}\pm 0.6$ & 6.9\\
      \BtoEtaMuMu   &      &  & $-2.0^{+9.7}_{-6.6}\pm 0.4$ & 11.2 \\%%$-2.0^{+9.7}_{-6.6}\pm 0.4$\hspace{0.7cm}  & 15.9\\
      \hspace{1em}\BtoEtaGGMuMu   & 0.102 & $-0.4^{+1.7}_{-1.3}$ &  & \\ %\multr{2}{*}{$-2.0^{+9.7}_{-6.6}\pm 0.4$} & \multr{2}{*}{11.2}\\
      \hspace{1em}\BtoEtaThPiMuMu & 0.063 & $-0.1^{+0.7}_{-0.4}$ &  & \vspace{0.2cm}\\ 
      \B\to\piee    & & & $4.0^{+5.1}_{-4.2}\pm 1.6$   & 11.0\\
      \B\to\pimm    & & & $-0.9^{+3.9}_{-3.0}\pm 1.2$  & 5.0\vspace{0.2cm}\\
      \Bp\to\pipll  & & & $2.5^{+3.9}_{-3.3}\pm 1.2$   & 6.6\\
      \Bz\to\pizll  & & & $1.2^{+3.9}_{-3.3}\pm 0.2$   & 5.3\\
      \Bz\to\etall  & & & $-2.8^{+6.6}_{-5.2}\pm 0.3$  & 6.4\vspace{0.2cm}\\
      \B\to\pill    & & & $2.5^{+3.3}_{-3.0}\pm 1.0$   & 5.9\\\hline\hline
    \end{tabular}
  \end{center}
  \label{tab:pillResults}
\end{table*}
We extract  branching fractions by  fitting the data  set with
the fit  model described in  Section \ref{sec:fit_model}.  Projections
of the PDFs and data sets in \mes and \DeltaE are shown for the isospin
averaged     \B\to\piee\    fit,     the     lepton-flavor    averaged
\Bz\to\etall\  fit,   and  the  isospin   and  lepton-flavor  averaged
\B\to\pill\  fit   in  Figs.\ 3-5,  respectively.   Figure\ 6  shows
$-2\ln\lambda$ as a function of  the branching fraction for the
\pill,   \pipll, \pizll\, \etall, \etaee, and \etamm   measurements.   Branching   fraction
measurements and  upper limits at 90\%  CL are given in  Table III for
each mode.
%%%%%%%%%%%%%%%%%%%%%%%%
%% Isospin Avg Figure %%
%%%%%%%%%%%%%%%%%%%%%%%% 
\begin{figure*}[!h]	
  \begin{center}	
    \includegraphics[width=0.775\textwidth]{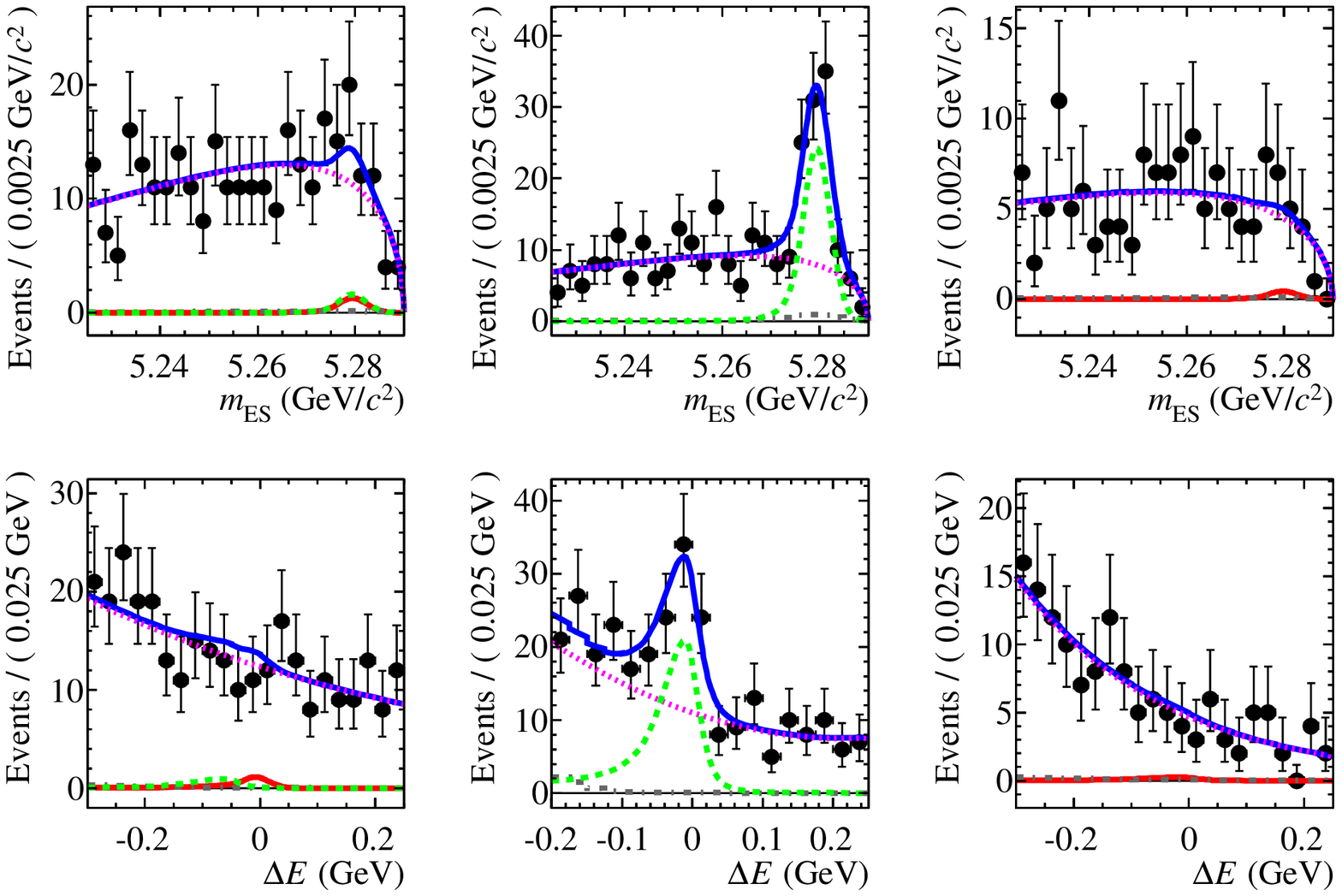}
    \setlength{\unitlength}{1in}
    \begin{picture}(0,0)
      \put(-4.950,3.60){\BtoPiEE}
      \put(-3.1525,3.60){\Bp\to\Kee}
      \put(-1.325,3.60){\BtoPiZEE}
      \put(-4.80,3.35)  {{\bf (a)}}
      \put(-3.0025,3.35){{\bf (c)}}
      \put(-1.175,3.35) {{\bf (e)}}
      \put(-4.80,1.5)   {{\bf (b)}}
      \put(-3.0025,1.5) {{\bf (d)}}
      \put(-1.175,1.5)  {{\bf (f)}}
    \end{picture}
    \caption{Fit projections  of \mes (top) and \DeltaE (bottom) for
      the isospin averaged \B\to\piee\  fit to the \pipee\ ((a) and (b)), 
      \Kee\ ((c) and (d)), and \pizee\ ((e) and (f)) data sets.
      Points with error bars represent data.  
      The  curves are  dotted (magenta)  for  combinatoric background,
      dot-dashed  (gray) for  $\Kstar/\KS\ellell$  background, dashed
      (green) for \Kll   signal  and   background,   and  solid   (red)
      for \pill\ signal.  The solid blue curve represents the total fit function. (color available online)}
  \end{center}	
  \label{fig:pieeDataFit}
\end{figure*}	

\begin{figure*}	
  \begin{center}	
    \includegraphics[width=0.775\textwidth]{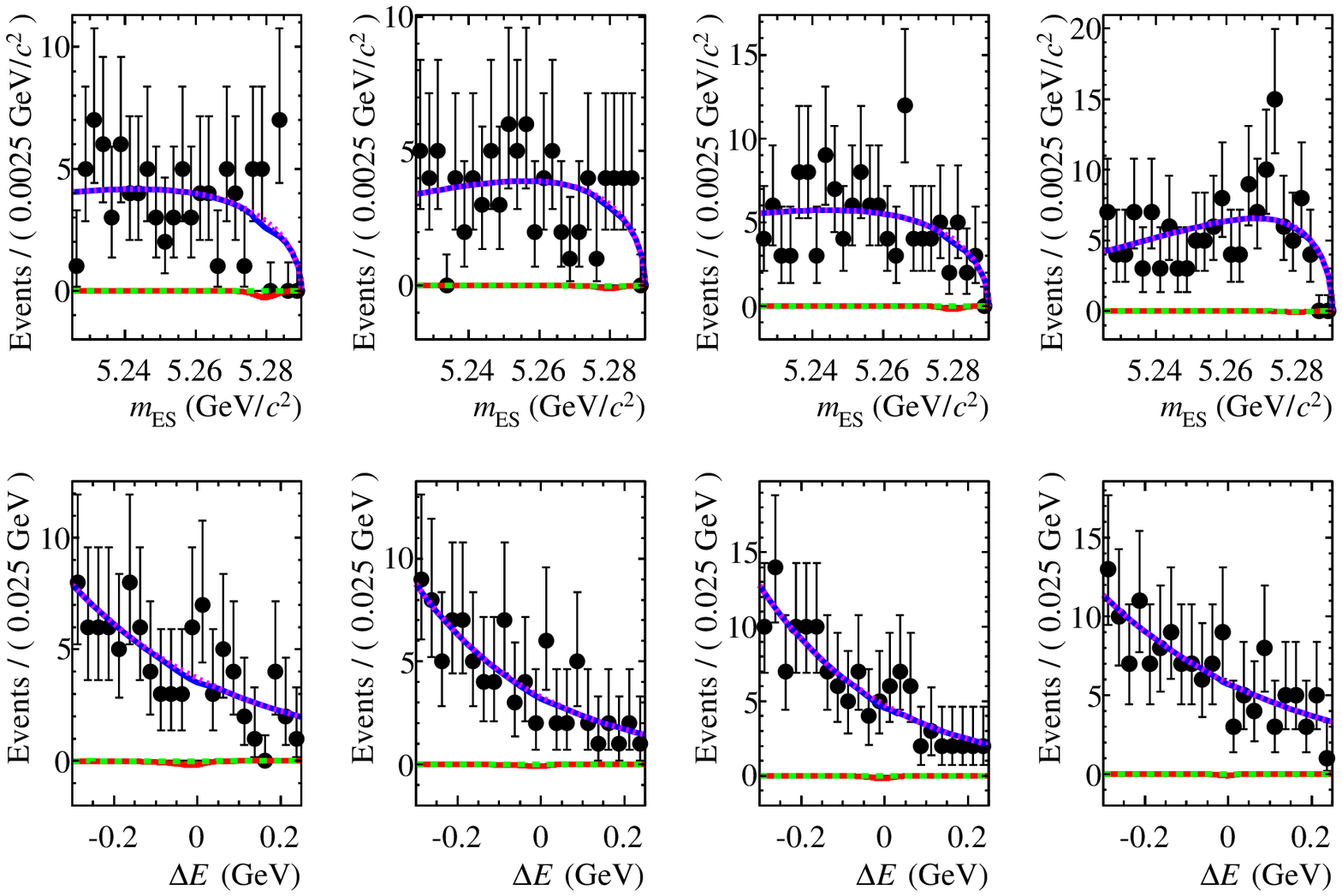}
    \setlength{\unitlength}{1in}
    \begin{picture}(0,0)
      \put(-5.155,3.60){\BtoEtaGGEE}
      \put(-3.800,3.60){\BtoEtaThPiEE}
      \put(-2.450,3.60){\BtoEtaGGMuMu}
      \put(-1.100,3.60){\BtoEtaThPiMuMu}
      \put(-5.000,3.375){{\bf (a)}}
      \put(-3.625,3.375){{\bf (c)}}
      \put(-2.250,3.375){{\bf (e)}}
      \put(-0.900,3.375){{\bf (g)}}
      \put(-5.000,1.525){{\bf (b)}}
      \put(-3.625,1.525){{\bf (d)}}
      \put(-2.250,1.525){{\bf (f)}}
      \put(-0.900,1.525){{\bf (h)}}
    \end{picture}
    \caption{Fit projections  of \mes  (top) and \DeltaE  (bottom) for
      the  lepton-flavor averaged  fit  to the  \etaggee\ ((a) and (b)),
      \etathpiee\  ((c) and (d)),  \etaggmm\  ((e) and (f)),  and
      \etathpimm\ ((g) and (h)) samples. Points with error bars represent data.
      The curves are dotted (magenta)
      for combinatoric  background, dashed (green) for self-crossfeed, and
      solid (red) for \etall\ signal.  The solid blue curve represents the total
      fit function. (color available online)}
  \end{center}	
  \label{fig:etallDataFit}
\end{figure*}	

%%%%%%%%%%%%%%%%%%%%%%%%%%%%%%%%%%%%%%%%%%
%% Lepton Flavor and Isospin Avg Figure %%
%%%%%%%%%%%%%%%%%%%%%%%%%%%%%%%%%%%%%%%%%%
\begin{figure*}	
  \begin{center}	
    \includegraphics[width=\textwidth]{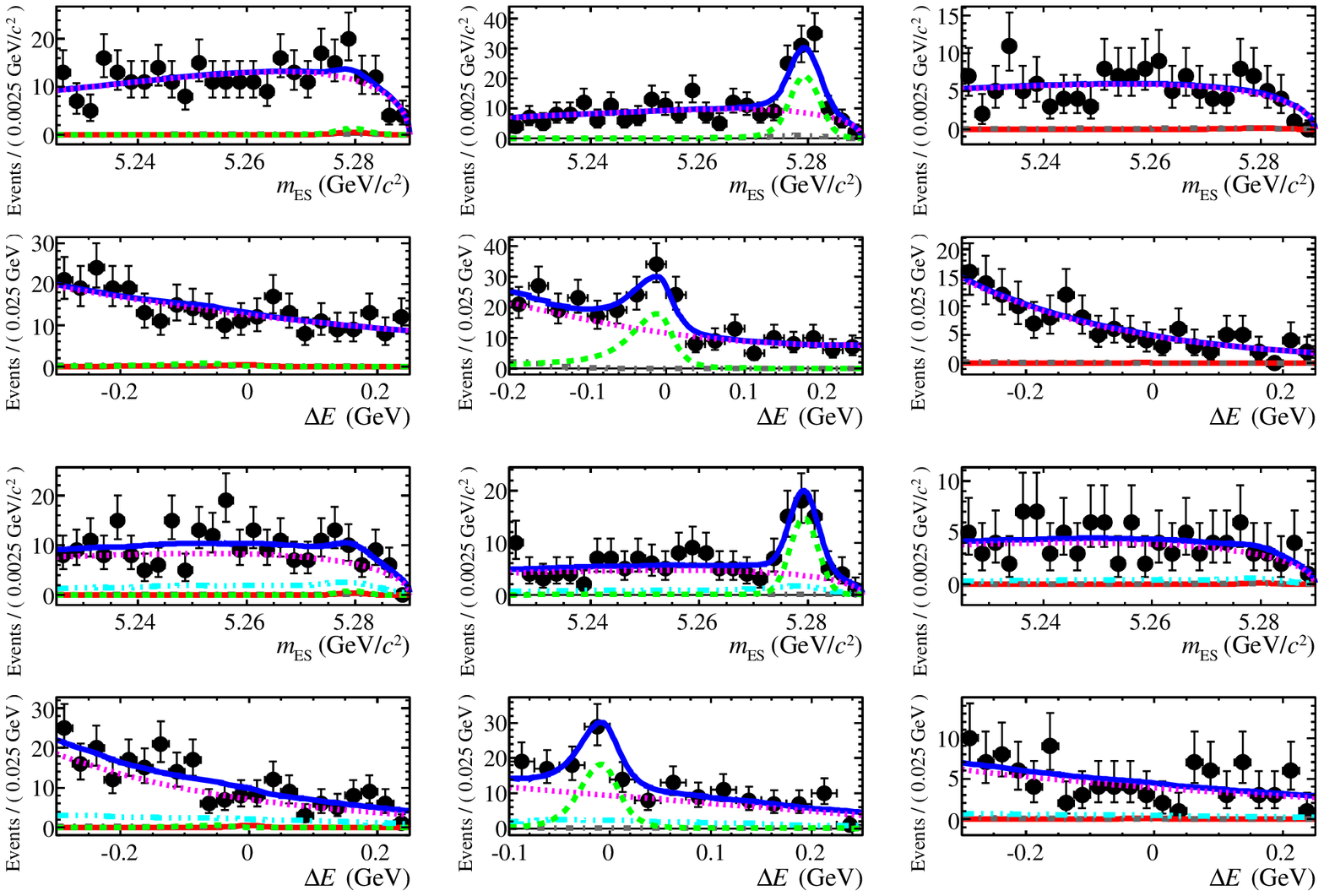}
    \setlength{\unitlength}{1in}
    \begin{picture}(0,0)
      \put(-2.700,4.92){\BtoPiEE}
      \put(-0.350,4.92){\Bp\to\Kee}
      \put(2.000,4.92){\BtoPiZEE}
      \put(-2.700,2.525){\BtoPiMuMu}
      \put(-0.350,2.525){\Bp\to\Kmm}
      \put(2.000,2.525){\BtoPiZMuMu}
      \put(-2.350,4.725){{\bf (a)}}
      \put(0.000,4.725){{\bf (b)}}
      \put(2.365,4.725){{\bf (c)}}
      \put(-2.350,3.525){{\bf (d)}}
      \put(0.00,3.525){{\bf (e)}}
      \put(2.365,3.525){{\bf (f)}}
      \put(-2.350,2.33){{\bf (g)}}
      \put(0.00,2.33){{\bf (h)}}
      \put(2.365,2.33){{\bf (i)}}
      \put(-2.350,1.13){{\bf (j)}}
      \put(0.00,1.13){{\bf (k)}}
      \put(2.365,1.13){{\bf (l)}}
    \end{picture}
    \caption{Fit  projections  of \mes and
      \DeltaE for  the  lepton flavor  and
      isospin  averaged \pill\  fit  to the \pipee\ ((a) and (d)), 
      \Kee\ ((b) and (e)), \pizee\ ((c) and (f)), \pipmm\ ((g) and (j)), 
      \Kmm\ ((h) and (k)), and \pizmm\ ((i) and (l)) data sets.  Points
      with error bars represent data.  The  curves are  dotted
      (magenta) for combinatoric  background, triple dot-dashed (cyan)
      for hadronic  peaking background in the  \mumu modes, dot-dashed
      (gray) for $\Kstar/\KS\ellell$  background, dashed (green) for \Kll
      signal  and background,  and solid  (red) for \pill\  signal.  The solid
      blue curve represents the total fit function. (color
      available online)}
  \end{center}	
  \label{fig:pillDataFit}
\end{figure*}	

As a  cross-check, we measure  the \Bp\to\Kll branching  fractions and
find    them   consistent    with   the    current    world   averages
\cite{Nakamura:2010zzi}.

We  set upper limits  on the  lepton-flavor  averaged branching
fractions of
%\begin{linenomath}
  \begin{eqnarray}
    \BR(\Bp\to\pipll) & < & 6.6  \times 10^{-8},\\
    \BR(\Bz\to\pizll) & < & 5.3  \times 10^{-8},\\
    \BR(\Bz\to\etall) & < & 6.4  \times 10^{-8},
  \end{eqnarray}
%\end{linenomath}
all at  the 90\% CL.   A lepton-flavor and isospin  averaged branching
fraction upper limit of
%\begin{linenomath}
  \begin{equation}
    \BR(B\to\pill) < 5.9 \times 10^{-8}
  \end{equation}
%\end{linenomath}
is set at 90\% CL.
Branching fraction measurements and upper limits at 90\% CL for the modes
\BtoPiEE, \BtoPiZEE, \BtoPiMuMu, \BtoPiZMuMu, \BtoEtaEE, and \BtoEtaMuMu are
listed in Table III.
%%%%%%%%%%%%%%%%%%%%%%%%
%% pi l+l- NLL Figure %%
%%%%%%%%%%%%%%%%%%%%%%%%
\begin{figure*}
  \begin{center}
   \includegraphics[width=0.85\textwidth]{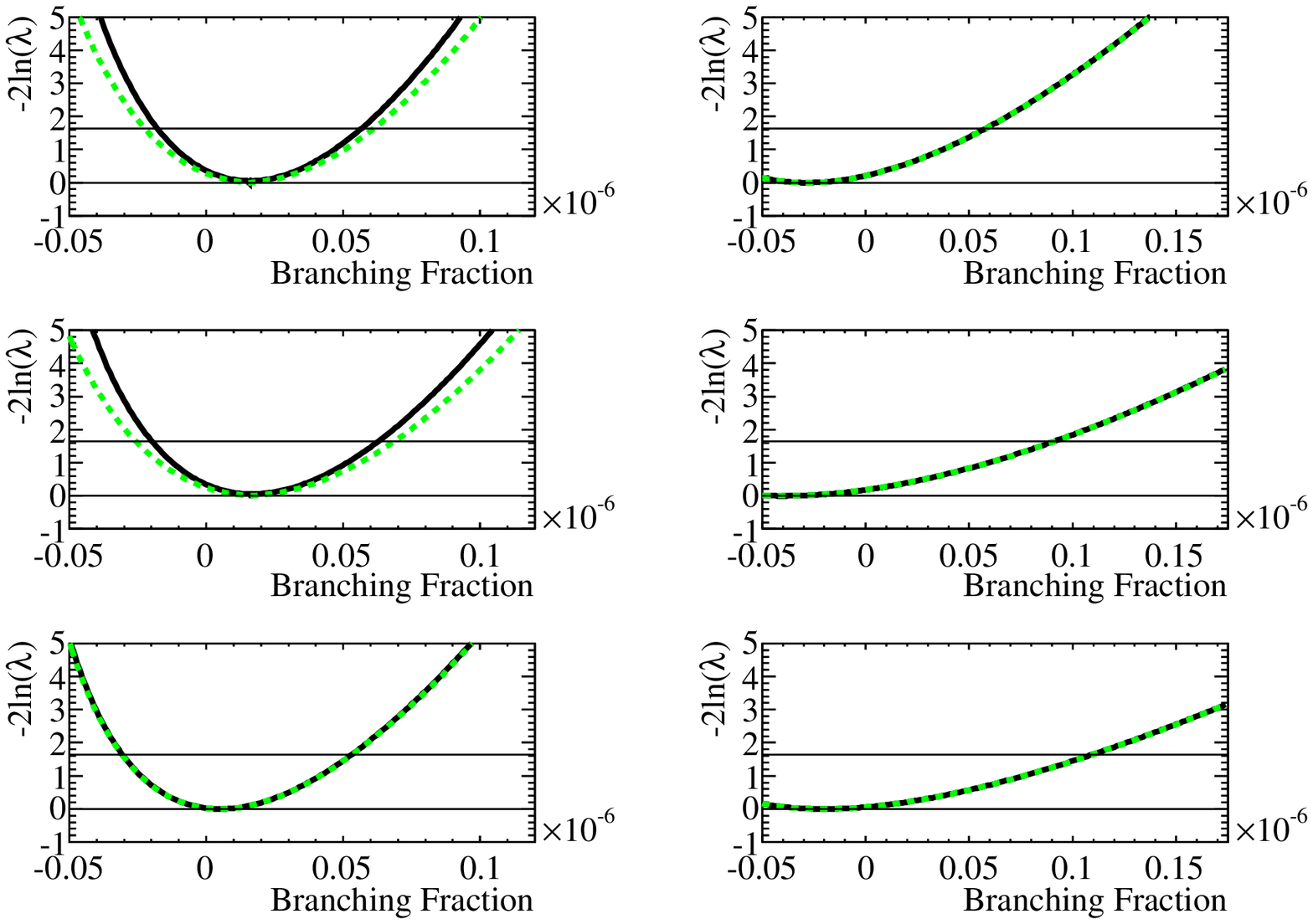}
  \setlength{\unitlength}{1in}
    \begin{picture}(0,0)
      \put(-5.150,3.700){{\bf (a)} \B\to\pill}
      \put(-5.150,2.350){{\bf (c)} \Bp\to\pipll}
      \put(-5.150,1.000){{\bf (e)} \Bz\to\pizll}
      \put(-2.300,3.700){{\bf (b)} \Bz\to\etall}
      \put(-2.300,2.350){{\bf (d)} \Bz\to\etaee}
      \put(-2.300,1.000){{\bf (f)} \Bz\to\etamm}
    \end{picture}
   \caption{The negative logarithm of the profile likelihood as a
      function    of     branching    fraction    for  (a) $\B\to\pill$, (b) $\Bz\to\etall$,
      (c) $\Bp\to\pipll$, (d) $\Bz\to\etaee$, (e) $\Bz\to\pizll$, and (f) $\Bz\to\etamm$
      The solid, black lines  are  the negative  log
      likelihood  curves including  only statistical  errors,  and the
      dashed, green  lines correspond to  the same  curves convolved  with a
      Gaussian distribution whose width is  equal to  the total  systematic error.
      (color available online)}
  \end{center}
  \label{fig:pillNLLCurves}
\end{figure*}

In conclusion,  we have searched  for the rare decays  $B\to\pill$ and
$\Bz\to\etall$ in  a sample of 471  million \BB decays  and observe no
statistically significant signal in any of the decay channels studied.
We set a lepton-flavor and isospin averaged upper limit at the 90\% CL
of $\BR(B\to\pill)<5.9\times 10^{-8}$, within a factor of three of the
SM expectation.   We also set  lepton-flavor averaged upper  limits of
$\BR(\Bp\to\pipll) < 6.6 \times  10^{-8}$ and $\BR(\Bz\to\pizll) < 5.3
\times 10^{-8}$.  Branching fraction upper limits at 90\% CL have also been calculated for
the modes \BtoPiEE, \BtoPiZEE, \BtoPiMuMu, and \BtoPiZMuMu.  Our upper limits on
the \BtoPiZEE, \BtoPiZMuMu, and \Bz\to\pizll\ branching fractions are the lowest upper limits to date.
The results presented for the \pill\ modes supersede those of the previous \babar\
analysis~\cite{Aubert:2007mm}.
We   have   also  performed   the   first   search   for  the   decays
\Bz\to\etall and set an  upper limit  on the  lepton-flavor averaged
branching  fraction of  $\BR(\Bz\to\etall)<6.4\times  10^{-8}$ at  the
90\% CL.  Upper limits at 90\% CL for the \BtoEtaEE and
\BtoEtaMuMu branching fractions have been reported.  

%% file: acknowledgements.tex
We are grateful for the 
extraordinary contributions of our \pep2\ colleagues in
achieving the excellent luminosity and machine conditions
that have made this work possible.
The success of this project also relies critically on the 
expertise and dedication of the computing organizations that 
support \babar.
The collaborating institutions wish to thank 
SLAC for its support and the kind hospitality extended to them. 
This work is supported by the
US Department of Energy
and National Science Foundation, the
Natural Sciences and Engineering Research Council (Canada),
the Commissariat \`a l'Energie Atomique and
Institut National de Physique Nucl\'eaire et de Physique des Particules
(France), the
Bundesministerium f\"ur Bildung und Forschung and
Deutsche Forschungsgemeinschaft
(Germany), the
Istituto Nazionale di Fisica Nucleare (Italy),
the Foundation for Fundamental Research on Matter (The Netherlands),
the Research Council of Norway, the
Ministry of Education and Science of the Russian Federation, 
Ministerio de Econom\'{\i}a y Competitividad (Spain), and the
Science and Technology Facilities Council (United Kingdom).
Individuals have received support from 
the Marie-Curie IEF program (European Union) and the A. P. Sloan Foundation (USA).

%% file: btodll_prd.bbl
\begin{thebibliography}{99}
\bibitem{legendremom}
The momentum weighted polynomials are defined as  $L^j_i   =
\sum^N_{k=0} |\vec{p}_k|^j\cos^i\theta_k$ with  the sum running over
$N$  particles (charged tracks and EMC clusters) in  the event.   The  angle $\theta_k$  is the  angle
between  $\vec{p}_k$, the  momentum of  the $k$-th  particle  in the
event, and some predefined direction, \eg\ the thrust axis, the
ROE thrust  axis, or  the dilepton momentum.
\end{thebibliography}
